\documentclass[11pt]{article}

\usepackage{pdfpages}

\usepackage[utf8]{inputenc}

\usepackage{lineno}
\usepackage[numbers]{natbib}
\usepackage{color}
\usepackage{graphicx}
\usepackage{amsmath}
\usepackage{amssymb}
\usepackage{float}
\usepackage{authblk}
\usepackage{geometry}
\usepackage{indentfirst}
\usepackage[colorlinks=true,allcolors=blue]{hyperref}
\usepackage[english]{babel}
\numberwithin{equation}{section}

\definecolor{qiblue}{rgb}{0, 0, 1}
\definecolor{gred}{rgb}{1, 0, 0}
\definecolor{jbpgreen}{rgb}{.1, .8, .1}

\usepackage[english]{babel}
\begin{document}





\title{The arrow of evolution when the offspring variance is large}

\author[1]{Guocheng Wang}
\author[2,3,5]{Qi Su}
\author[1,4]{Long Wang}
\author[2,3,5]{Joshua B.~Plotkin}

\affil[1]{\footnotesize Center for Systems and Control, College of Engineering, Peking University, Beijing 100871, China}
\affil[2]{Center for Mathematical Biology, University of Pennsylvania, Philadelphia, PA 19104, USA}
\affil[3]{Department of Mathematics, University of Pennsylvania, Philadelphia, PA 19104, USA}
\affil[4]{Center for Multi-Agent Research, Institute for Artificial Intelligence, Peking University, Beijing 100871, China}
\affil[5]{Department of Biology, University of Pennsylvania, Philadelphia, PA 19104, USA}

\maketitle

\begin{abstract}
\noindent The concept of fitness is central to evolution, but it quantifies only the expected number of offspring an individual will produce. The actual number of offspring is also subject to noise, arising from environmental or demographic stochasticity. 
In nature, individuals who are more fecund tend to have greater variance in their offspring number -- sometimes far greater than the Poisson variance assumed in classical models of population genetics. Here, we develop a model for the evolution of two types reproducing in a population of non-constant size. The frequency-dependent fitness of each type is determined by pairwise interactions in a prisoner's dilemma game, but the offspring number is subject to an exogenously controlled variance that may depend upon the mean. 
Whereas defectors are preferred by natural selection in classical well-mixed populations, since they always have greater fitness than cooperators, we show that large offspring variance can reverse the direction of evolution and favor cooperation. 
Reproductive over-dispersion produces qualitatively new dynamics for other types of social interactions, as well, which cannot arise in populations with a fixed size or Poisson offspring variance.
\end{abstract}

\clearpage
\section{Introduction}

The past decades have seen a proliferation of research using evolutionary theory to study social traits, in the fields of biology, animal behavior, and even social science \cite{Nowak1998,Nowak2004,Hilbe2018a,Ohtsuki2006,Weitz2016,Tilman2020,Allen2017a}.
Most of this theoretical development has been based on mathematical models that assume either infinite populations \cite{Taylor1978,Schuster1983,Nowak2006a,Weitz2016,Tilman2020}  or finite populations of constant size \cite{Hilbe2018a,Ohtsuki2006,Nowak2004}.
Despite these simplifying assumptions, mathematical models provide rich insights into how exogenous and intrinsic factors drive evolutionary dynamics of social behavior. 
For example, 
the literature has produced a rich set of explanations for cooperation based on repeated interactions, the establishment of reputations, and various forms of population structure \cite{Nowak1992,Ohtsuki2006,Tarnita2009,Allen2017a,McAvoy2021,Su2022,Su2022nhb,Cooney2019,Hilbe2018a,Santos2018,Nowak2005,Nowak1993,Nowak2006fiverule,Stewart2013}. Several key theoretical insights have been validated by controlled experiments on human subjects \cite{Gachter2009,Yamauchi2011,Yoelia2013,Greiner2005}. This field of research has been so successful that the question of how cooperation can be favored by natural selection, famously posed by Darwin, is now not only resolved, but resolved in several distinct ways applicable in different contexts.

Here we reveal an qualitatively different and pervasive mechanism that can promote cooperation by natural selection or payoff-biased imitation. 
Most mechanisms known to support cooperation boil down to some form of population structure \cite{Kay2020} -- either physical limitations on social interactions, reproduction, or imitation, or structure imposed by tags or reputations. By contrast, here we describe a much more simple scenario that can favor cooperation in a population that lacks any form of exogenous or endogenous structure. We show that demographic stochasticity, which is \textit{a priori} a realistic feature of any natural population, can by itself promote social behaviors that would otherwise be suppressed in idealized populations of constant (or infinite) size.

There is precedent for the idea that demographic stochasticity alters evolutionary dynamics. The fact that mortality, reproduction, and migration are subject to demographic fluctuations in populations -- as well as processes of imitation and innovation -- is known influence the dynamics of competing types under frequency-independent selection \cite{Parsons2007,Parsons2007a,Parsons2010,McKane2005,Butler2009,Hallatschek2007,Stollmeier2018,Wienand2017,Taitelbaum2020,Chotibut2017} and also frequency-dependent selection \cite{Constable2016,Houchmandzadeh2012,Houchmandzadeh2015,Huang2015}.
For example, when a population contains two types with the same expected number of offspring, one type can be favored when the population size is small, and the other type favored when the population size is near to its carrying capacity \cite{Parsons2007,Parsons2007a,Parsons2010}.
And a few studies have shown that demographic stochasticity can even reverse the direction of natural selection, promoting a type that would otherwise be disfavored without stochasticity  \cite{Constable2016,Houchmandzadeh2012,Houchmandzadeh2015}.

Nonetheless, prior work on selection with demographic stochasticity has either assumed constant fitness, in which one's fitness is independent of the composition of the population, or assumed different carrying capacities for different phenotypes, e.g.,~producers enjoy a larger carrying capacity than non-producers \cite{Constable2016,Houchmandzadeh2012,Houchmandzadeh2015}.
Most models of demographic stochasticity 
also assume that offspring numbers follow a Poisson distribution \cite{Constable2016,Huang2015,Parsons2010,Parsons2007,Parsons2007a,Houchmandzadeh2012,Houchmandzadeh2015}, so that the mean and variance in offspring number are identical. But empirical field studies have found that over-dispersion in offspring number (variance exceeding mean) is commonplace across diverse taxa \cite{Zuur2009,Linden2011,VerHoef2007,Richards2008}.

In this paper, we develop a general framework to study evolutionary dynamics with demographic stochasticity, which can capture both frequency-dependent fitness, arising from social interactions, as well as over-dispersion in the number of offspring. We provide a simple analytical condition that governs the long-term outcome of competition between multiple types. Applied to pairwise social interactions involving cooperation or defection, we find that demographic stochasticity can favor cooperators provided the offspring variance is sufficiently large, even without any other mechanisms. For more general pairwise payoff structures, we show that demographic stochasticity can reverse the stability of equilibria, from coexistence to bi-stability and vice versa, or from dominance of one type to dominance of another. Our analysis highlights the profound effects of demographic stochasticity on the evolution of interacting types in a population.

\begin{figure}[t]
    \centering
    \includegraphics[width=\textwidth]{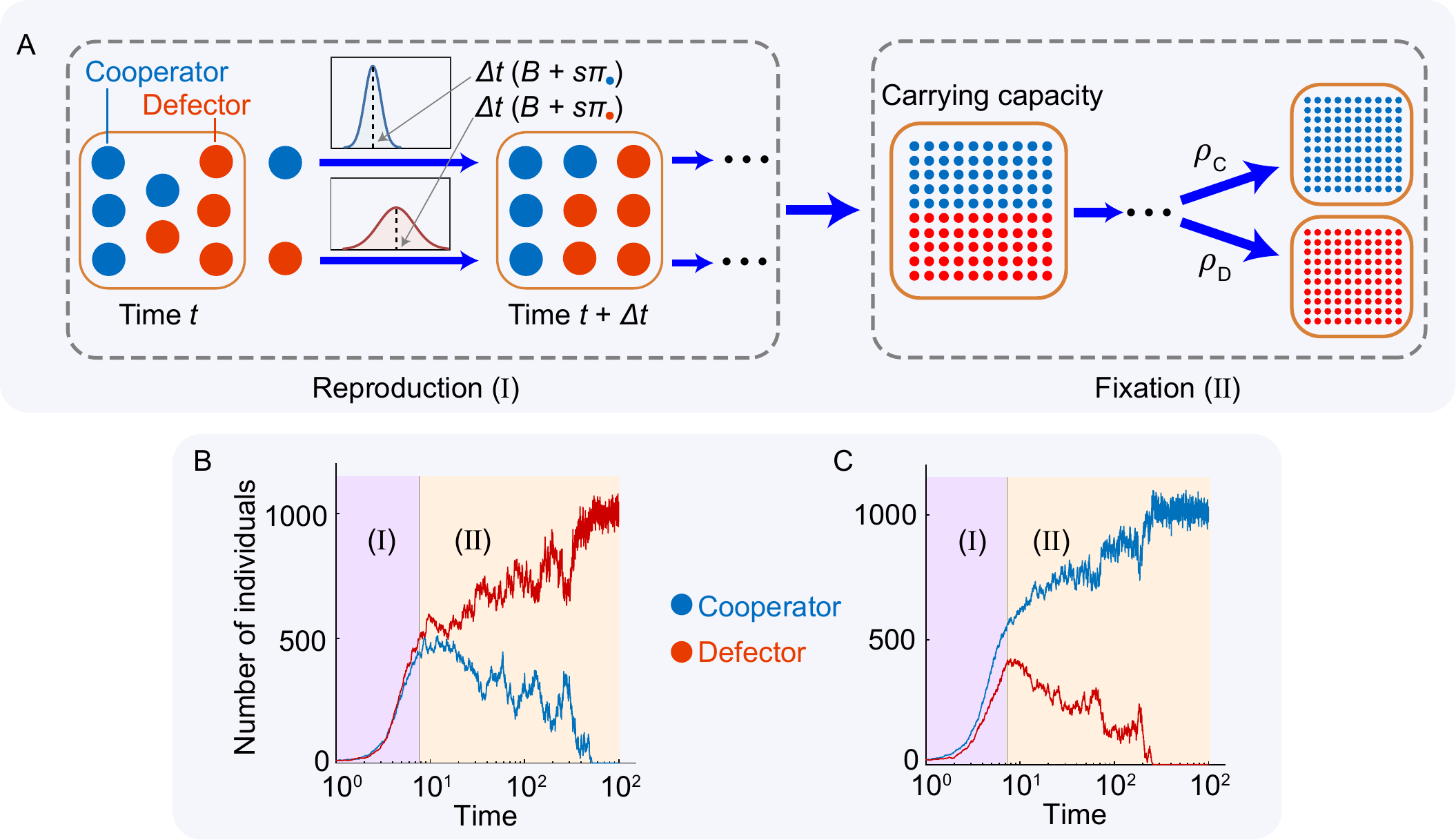}
    \caption{\textbf{Evolutionary dynamics with demographic stochasticity.} 
    (A) Competition between cooperators (blue circle) and defectors (red circle) in a stochastic population of non-constant size.  Each individual $i$ derives payoff $\pi_i$ from pairwise game-play with each other individual in the population. The number of offspring produced by an individual within time $\Delta t$ has mean $(B+s\pi_i)\Delta t$ and variance $(\delta_1B+\delta_2 s\pi_i)\Delta t$, which are both higher for defectors than for cooperators. When selection is weak ($s \ll \alpha$), the population quickly reaches carrying capacity (during time period I) while the frequency of cooperators and defectors remains unchanged from its initial value ($p_0=1/2$ shown here). Thereafter (time period II) the population remains near carrying capacity ($M \approx 1000$ shown here), while the frequency of cooperators and defectors slowly vary until either cooperators go extinct (example in panel B) or defectors go extinct (panel C).
    Parameters: $b=3$, $c=1$, $s=0.01$, $\delta_1=\delta_2=1$, $x_0=y_0=10$, $\lambda=1\times10^{-3}$, $B=2$, $D=1$.}
    \label{fig1}
\end{figure}

\section{Model}

We first consider an evolving population of two types: cooperators (C) and defectors (D).
Each individual interacts pairwise with each other, in which the cooperator pays a cost $c$ to bring his opponent a benefit $b$ ($b>c$), and the defector pays no cost and provides no benefit. In other words, pairwise interactions follow a simple ``donation game", which provides a minimal model for studying the evolution of cooperation \cite{rapoport1965prisoner}.
Following all pairwise interactions, 
each individual obtains an average payoff that will determine their reproductive output (or, equivalently, the number of individuals who copy their type by social contagion). In a population with $x$ cooperators and $y$ defectors, the cooperator's payoff (denoted by $\pi_C$) and the defector's payoff (denoted by $\pi_D$) are

\begin{subequations}
\begin{align}
    \pi_C=&\frac{x}{x+y}b-c, \\ \pi_D=&\frac{x}{x+y}b.
\end{align}
\end{subequations}

In a classic Moran model, each birth event is followed by a death event, and so the population size remains constant. Here we remove this constraint by decoupling the birth and death events. Births are assumed to follow a continuous-time Markov process with independent and stationary increments (see Section S1 in Supplementary Information), such that the expected number of offspring individual $i$ produces per unit time is \begin{equation}
     \mathbb{E}(\xi_i)=B+s\pi_i,
     \label{eq:mean}
\end{equation}
where $B$ is a baseline number of offspring, $\pi_i$ is individual $i$'s payoff, and the parameter $s>0$ is the intensity of selection. Note that the baseline birth rate is the same for all individuals, regardless of type, and it does not depend upon payoffs from social interactions. The selection intensity $s$ measures to what degree the payoff derived from social interactions affects the offspring number. In this paper we focus on the case of weak selection ($s\ll 1$), a regime widely adopted in the literature \cite{Allen2017a,McAvoy2021,Nowak2004,Ohtsuki2006}. Since the defector's payoff $\pi_D$ is larger than the cooperator's payoff $\pi_C$ in any population state, defectors always have a greater expected fecundity (Fig.~\ref{fig1}). 

To fully describe the birth process, we also specify the variance in the number of offspring. We are particularly interested in cases of over-dispersion, which can be modelled in many alternative ways \cite{Linden2011,VerHoef2007}, such as a quasi-Poisson model (variance proportional to mean), mixed-effects Poisson model, and negative binomial model (variance a quadratic function of mean). Here we study a general class of Markov birth models by stipulating
\begin{equation}
    {\rm Var}(\xi_i)=\delta_1 B+\delta_2 s\pi_i,
    \label{eq:variance}
\end{equation}
where parameters $\delta_1$ and $\delta_2$ measure the magnitude of offspring variance ${\rm Var}(\xi_i)$ relative to the mean $\mathbb{E}(\xi_i)$. The parameter $\delta_1$ controls how  offspring variance is influenced by the baseline birth rate; and $\delta_2$ controls how  offspring variance is influenced by payoffs from social interactions. Specific choices of $\delta_1$ and $\delta_2$ reduce to well-known classical models, such as a deterministic system ($\delta_1=\delta_2=0$) or a Poison birth process ($\delta_1=\delta_2=1$). In the regime of weak selection, the number of offspring produced per unit time is over-dispersed whenever  $\delta_1>1$.

Death events are modelled as a Poisson process, arising from two rates that are summed. First, an individual dies at constant baseline rate, $D$. Second, in order to model competition for limited resources, additional deaths occur at rate $\lambda$ times the current total population size.

\section{Results}

\subsection{Evolution of cooperation with demographic stochasticity}
\label{section3.1}
Let $x$ and $y$ denote the number of cooperators and defectors respectively, which will change over time. 
Given the class of models described above for the payoff-dependent birth-process and the population-size dependent death process,  the evolutionary dynamics of $x$ and $y$ can be approximated by a two-dimensional It\^o stochastic differential equation (see Section S1 in Supplementary Information):
\begin{subequations}
    \begin{align}
    {\rm d}x&=x\left[\alpha+s\pi_{\rm C}-\lambda (x+y)\right] {\rm d}t+ \sqrt{x\left[\delta_1 B+\delta_2 s\pi_{\rm C}+D+\lambda(x+y)\right]}{\rm d}W^{(1)}_t, \label{eq:ito_diffusion_a} \\
    {\rm d}y&=y\left[\alpha+s\pi_{\rm D}-\lambda (x+y)\right]{\rm d}t+ \sqrt{y\left[\delta_1 B+\delta_2 s\pi_{\rm D}+D+\lambda(x+y) \right]}{\rm d}W^{(2)}_t, \label{eq:ito_diffusion_b} 
    \end{align} \label{eq:ito_diffusion} 
\end{subequations}
where $\alpha =B-D>0$ indicates the net growth rate from baseline birth and death events, and $W_t^{(1)}$ and $W_t^{(2)}$ are independent standard Wiener processes. Although the birth process can be over-dispersed in our model (when $\delta_1>1$), deaths follow a simple Poisson process with variance equal to mean. 

To study how the relative abundance of cooperators and the total population size evolve over time, we make the co-ordinate transformation $(p,n)=(x/(x+y),x+y)$. Applying It\^o's lemma in Eq.~\ref{eq:ito_diffusion}, the system can then be described by the equations
\begin{subequations}
\begin{align}
          {\rm d}p=&scp(1-p)\left(-1 +\frac{\delta_2}{n}\right){\rm d}t+\frac{y}{n^2} \sqrt{x(\delta_1 B+D+\lambda n) }{\rm d}W^{(1)}_t  \notag\\
          &-\frac{x}{n^2}\sqrt{y(\delta_1 B + D+\lambda n)}{\rm d}W^{(2)}_t, \label{eq:transformed_a}\\
          {\rm d}n=&[n\alpha+s(b-c)pn-\lambda n^2]{\rm d}t+ \sqrt{x(\delta_1 B +D+\lambda n)}{\rm d}W^{(1)}_t   \notag \\
          & +\sqrt{y(\delta_1 B +D+\lambda n)}{\rm d}W^{(2)}_t. \label{eq:transformed_b}
\end{align}
\label{eq:transformed}
\end{subequations}

The simple case in which stochasticity is absent (i.e., $\delta_1=\delta_2=0$ for births, and no variance for deaths) provides a deterministic reference point for comparison to any stochastic system. 
In the deterministic system, ${\rm d}p$ is always negative and the abundance of cooperators continuously decreases until cooperators reach extinction. Thus, cooperation is never favored by natural selection in the deterministic limit. Moreover, in this deterministic limit, changes in the total population size $n$ depend on both $p$ and $n$. But for sufficiently weak selection intensity ($s\ll \alpha$), changes in the total population size $n$ are much more rapid than changes in the cooperator frequency, $p$.
In the regime of weak selection, before $p$ changes its value at all, $n$ has grown logistically to its equilibrium value $(\alpha+s(b-c)p)/\lambda$, which we denote by $M$. $M$ is called carrying capacity, and it describes the maximum number of individuals that the environment can sustain.  When the net growth rate is much larger than selection intensity, $\alpha \gg s$, the carrying capacity is well approximated by $M\approx \alpha/\lambda$. 
\begin{figure}[t]
    \centering
    \includegraphics[scale=0.75]{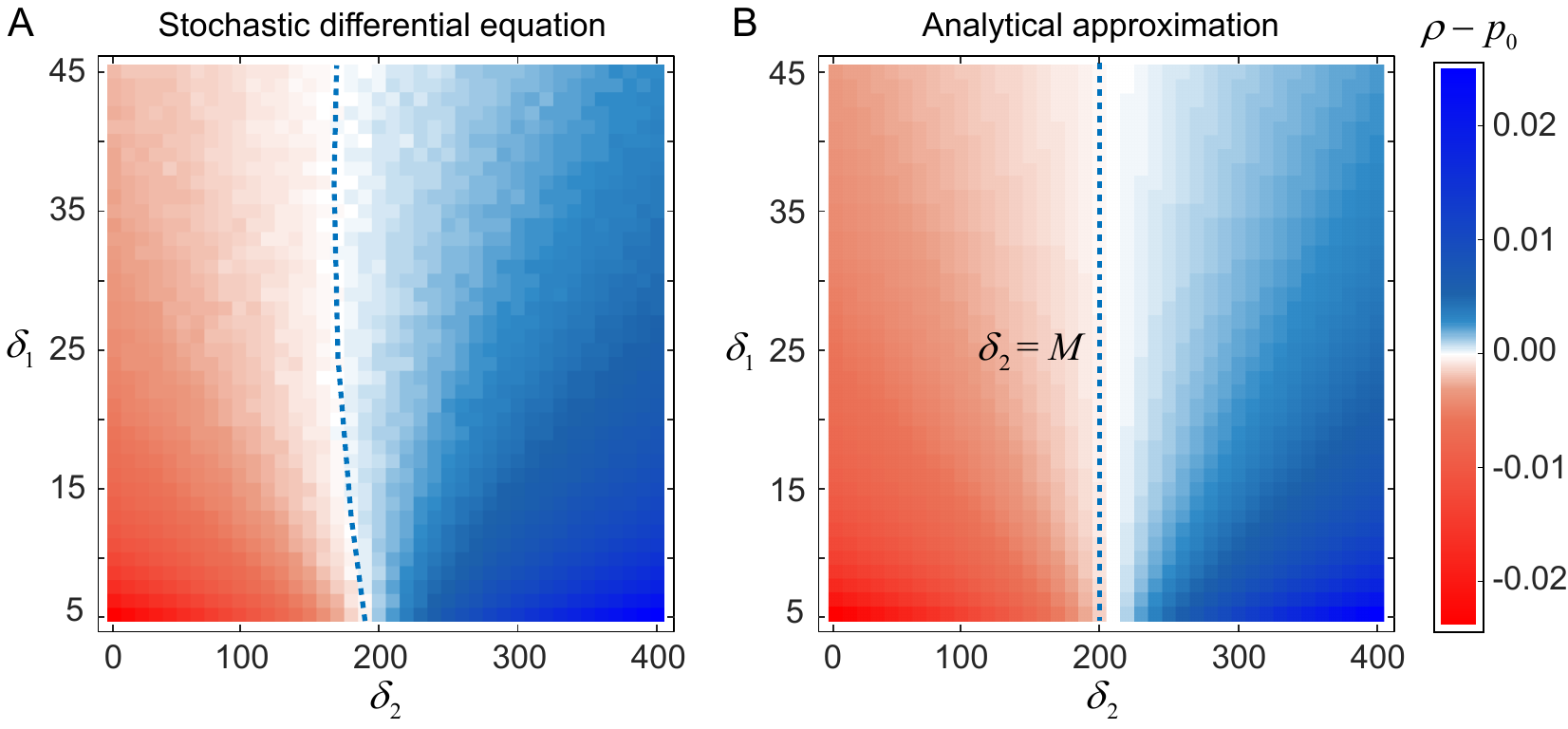}
    \caption{\textbf{Demographic stochasticity can favor the evolution of cooperation.} Colors represent the fixation probability of cooperation relative to neutral drift, $\rho - p_0$, as a function of parameters $\delta_2$ and $\delta_1$. We say that selection favors cooperation when cooperators are more likely to fix than under neutrality (blue regions). Panel (A) shows exact solutions sampled from the stochastic differential equation (Eq.~\ref{eq:ito_diffusion}), whereas panel (B) shows the analytical approximation in the regime of weak selection (Eq.~\ref{eq:fixation prob}). The dashed line indicates the separation between regimes that favor cooperation (blue) or favor defection (red).
    Parameters: $B=2$, $D=1$, $s=0.005$, $b=1.1$, $c=1$, $\lambda=5\times 10 ^{-3}$, $x_0=y_0=50$.}
    \label{fig2}
\end{figure}

For a stochastic system ($\delta_1 \ne 0$ and $\delta_2 \ne 0$) the trajectories of $p$ and $n$ are not determined by the initial conditions alone, but depend upon chance events. We quantify the evolutionary advantage of cooperators by studying the fixation probability -- namely, the chance of absorption into the full-cooperation state ($p=1$). 
Starting from $x_0$ cooperators and $y_0$ defectors initially (thus $p_0=x_0/(x_0+y_0)$ and $n_0=x_0+y_0$), the fixation probability, denoted by $\rho(x_0,y_0)$ or $\rho(p_0,n_0)$, is the probability that at some time $t$ defectors become extinct while cooperators still exist, that is $y(t)=0$ but $x(t)>0$ \cite{Czuppon2018}. 
In the regime $s \ll \alpha$ the fixation probability can be calculated by separating the time-scale of changes in $p$ versus changes in $n$ \cite{Parsons2017}. This analysis is tantamount to assuming that the total population size $n$ rapidly reaches its carrying capacity, while $p$ remains unchanged from $p_0$, and that subsequently $p$ evolves in one dimension while the population size remains near the slow manifold $n=M$ (see Fig.~\ref{fig1} and Supplementary Fig.~S3). Under this analysis, we can approximate the fixation probability by a simple expression (Section S2.1 in Supplementary Information) 
\begin{equation}
    \rho(p_0,n_0) \approx p_0+\frac{sc}{(\delta_1+1) B}\left(\delta_2-M\right)p_0(1-p_0).
    \label{eq:fixation prob}
\end{equation}
We performed numerical simulations, drawing sample paths from the full SDE system given by  Eq.~\ref{eq:ito_diffusion}, to verify the accuracy of this analytic approximation for the fixation probability  (Fig.~\ref{fig2}).

Note that fixation probability does not depend on the initial population size, but rather on the initial frequency of cooperators. In the absence of selection ($s=0$), the fixation probability equals the initial frequency of cooperators, $p_0$. And so we say that cooperation is favored by selection if the fixation probability exceeds $p_0$, which will occur whenever
\begin{equation}
    \delta_2>M. \label{eq:condition}
\end{equation}
This simple condition tells us when demographic stochasticity causes selection to favor cooperators, even though selection disfavors cooperation in a deterministic setting. In particular, demographic stochasticity can favor the fixation of cooperators when the offspring variance is sufficiently large -- in particular, when $\delta_2$ exceeds the carrying capacity $M$. What matters for the direction of selection, then, is the size of the offspring variance arising from payoffs in social interactions, relative to its mean.

We can gain some useful intuition for the forces that govern the fate of cooperators by considering the deterministic part of Eq.~\ref{eq:transformed_a}. The first term in this equation, $-scp(1-p)$, represents the deterministic contribution to the evolution of cooperator frequency, which always opposes cooperators. Whereas the second term in this equation, $\delta_2scp(1-p)/n$, arises from demographic stochasticity and it always favors cooperators. Whether or not cooperation is favored overall depends upon the balance between these two forces -- the deterministic force suppressing cooperation and demographic stochasticity that favors cooperation. For $\delta_2<M$, the deterministic disadvantage is the stronger force and cooperators are net disfavored (recall that $n$ rapidly reaches carrying capacity $n=M$ before cooperators change frequency). However, if $\delta_2>M$, the stochastic advantage matters more, so that cooperators are favored, which constitutes an evolutionary reversal compared to a classical model without demographic stochasticity.

Other model parameters, $s$, $c$, $p_0$, $\delta_1$ and $B$, do not produce a reversal in the direction of selection for cooperation, but they nonetheless influence the fixation probability. For example, increasing $\delta_1$ or increasing the baseline birth rate $B$ moves the fixation probability towards the neutral value, $p_0$. Moreover, in the regime where demographic stochasticity favors cooperation, $\delta_2>M$, the fixation probability is increased yet further when the selection intensity $s$ is large or when the cost of cooperation $c$ is large (Eq.~\ref{eq:fixation prob}). Both of these results contravene the classical intuition that selection and the cost of cooperation should disfavor cooperators. 
We have performed simulations to verify the effects of all these parameters, in comparison to the analytical approximation (Supplementary Fig.~S2).

\subsection{An explicit birth-death process}

Our model of demographic stochasticity is quite general, stipulating only several properties of the Markov birth and death processes for competing types. We have analyzed this class of models by approximation, using a stochastic differential equation. In this section we construct an explicit example of birth and death processes that satisfy our model stipulations, and we compare the predictions of our SDE analysis to individual-based simulations of the discrete stochastic process.

Most prior studies of demographic stochasticity are based on a reproduction process with a single offspring per birth event, which naturally leads to a Poisson birth process \cite{Feller1950,Huang2015,Constable2016,Parsons2010,Parsons2007,Parsons2007a}. The Poisson process occurs as a special case within our family of models, when $\delta_1=\delta_2=1$. In this case, our analysis shows that demographic stochasticity alone cannot favor cooperation, because $\delta_2<M$. 
We will therefore consider non-Poisson birth process, in which the offspring produced per unit time is over-dispersed. This is a realistic scenario for many species, especially pelagic organisms, that have heavy-tailed offspring distributions \cite{Davis1956,Eldon2006}; as well as for social contagion \cite{brady2017emotion,schroder2021social}.

We will define a birth process by two factors: the times of birth events and the litter size (offspring number) in each such birth event. A natural way to describe this is through a compound Poisson process \cite{Last2017}. Specifically, for individual $i$ with payoff $\pi_i$, the times of birth events obey a Poisson process with intensity $\theta_i$. In each such birth event, the number of offspring produced (litter size) is also stochastic. We consider two cases: the litter size itself follows a Poisson distribution with mean $\mu_i$, or the litter size follows a negative binomial distribution with parameters $q_i$ and $m$ ($q_i\in [0,1]$ and $m\in \mathbb{N}^*$). Both of these distributions have been used to model litter sizes in empirical studies \cite{Sheldon2003,Richards2008,Linden2011}.

\begin{figure}[t]
    \centering
    \includegraphics[scale=0.75]{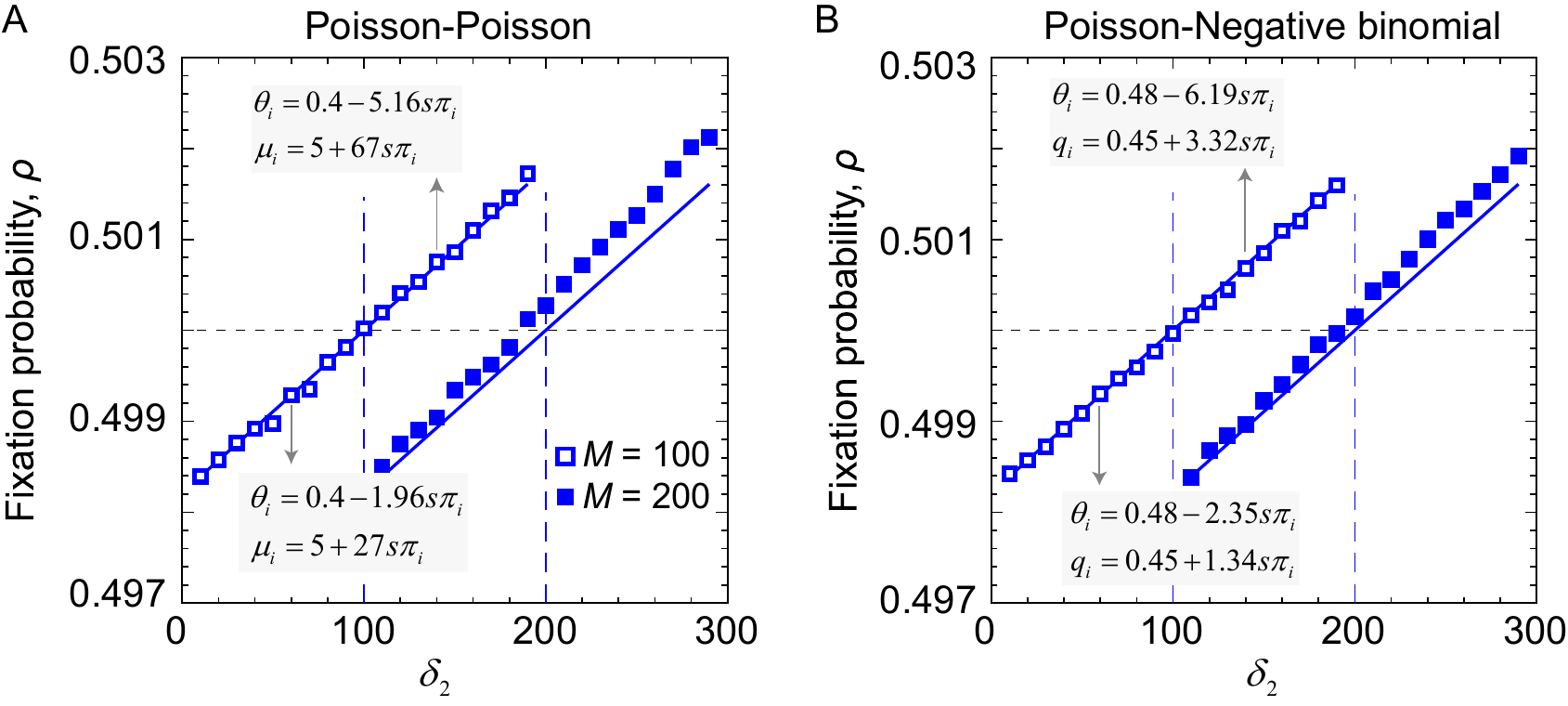}
    \caption{\textbf{Selection for cooperation in a compound Poisson birth process.}
    We simulated a compound Poisson birth process with either a Poisson-distributed litter size (A) or a negative binomial litter size (B). The parameters of the birth process ($\theta_i$ and $\mu_i$ in panel A; $\theta_i$ and $q_i$ in panel B) can be chosen to satisfy our general conditions for the mean and variance in total offspring produced per unit time, for any choice of $\delta_1>1$, $\delta_2$, and $B$.  
    Two examples with the parameters that correspond to $(\delta_1=6,\delta_2=60)$ and $(\delta_1=6,\delta_2=140)$ are shown in each panel. Blue squares indicate the fixation probability of cooperators, starting from an initial population with $x_0=y_0=50$, observed in $5\times 10^7$ replicate Monte Carlo simulations, with carrying capacity either $M=100$ or $M=200$. Selection favors cooperation if the fixation probability $\rho$ exceeds the initial fraction of cooperators,  0.5 (horizontal dashed line).
    The solid lines plot our analytical approximation for the fixation probability (Eq.~\ref{eq:fixation prob}). As predicted by our analysis, cooperation is favored when $\delta_2>M$.
    Parameters: $B=2$, $D=1$, $\delta_1=6$, $s=0.001$, $b=1.1$, $c=1$, $m=5$ (negative binomial), $x_0=y_0=50$, $\lambda=1/100$ ($M=100$) or $\lambda=1/200$ ($M=200$).}
    \label{fig3}
\end{figure}

The parameters of the compound Poisson process depend upon an individual's payoff and the selection intensity. For the Poisson-Poisson case (the litter size follows a Poisson distribution) the reproductive process of individual $i$ is characterized by parameters $\theta_i$ and $\mu_i$, and we assume that the payoff $\pi_i$ affects both $\theta_i$ and $\mu_i$ linearly
\begin{subequations}
\begin{align}
    \theta_i&=\theta_0+k_\theta s\pi_i, \\
    \mu_i&=\mu_0+k_\mu s\pi_i.
\end{align}
\end{subequations}
For the Poisson-negative binomial case (the litter size follows a negative binomial distribution), we assume that all individuals share the same $m$ and that payoffs affect $q_i$ and $\theta_i$ as follows:
\begin{subequations}
\begin{align}
    \theta_i&=\theta_0+k_\theta s\pi_i, \\
    q_i&=q_0+k_q s\pi_i.
\end{align}
\end{subequations}

Given these equations, we can always choose parameters of the compound Poisson process 
that satisfy our general stipulations on the mean and variance in the total offspring produced per unit time (Eq.~\ref{eq:mean} and Eq.~\ref{eq:variance}), provided $\delta_1>1$ and $\delta_2>0$ (see Section S3 in Supplementary Information). Note that for both of these compound Poisson birth processes (Poisson-Poisson and Poisson-Negative-Binomial) the total number of offspring produced per unit time must be over-dispersed ($\delta_1>1$).

We can compare Monte-Carlo simulations of these explicit population processes (discrete state, continuous time) to the analytical prediction for the fixation probability that we derived from a stochastic differential equation (Eq.~\ref{eq:fixation prob}). We find good agreement between the individual-based simulations and analytic approximations, for carrying capacities as small as $M=100$ or $M=200$  (Fig.~\ref{fig3}). Note that in both cases shown in Fig.~\ref{fig3}, for sufficiently large $\delta_2$ we have $k_\theta<0$ and $k_m>0$ or $k_q>0$. In other words, higher payoffs reduce the rate of birth events but increase the mean litter size per birth event; and when these effects are strong enough, then selection favors cooperation.

\subsection{Intuition for the effects of demographic stochasticity}

There is a simple intuition for how demographic stochasticity can favor cooperation in our class of models, even though cooperation is always disfavored in models with constant (or infinite) population size. The key insight has to do with the rapid growth of the total population size to carrying capacity, followed by slow dynamics in the frequency of cooperators near the manifold $n=M$. Importantly, during the slow dynamics there are still small fluctuations that move the population off the manifold $n=M$, followed by a rapid return back to carrying capacity. These small fluctuations have the effect of inducing an advective force pushing the frequency of cooperators $p$ in one direction or another. 

To be more precise, we have already noted that the total population size $n$ equilibrates much more quickly than the frequency of cooperators $p$ (Eq.~\ref{eq:transformed}), in the regime we study $\alpha \gg s$. And so, given an arbitrary initial state $p_0$ and $n_0$, $n$ will quickly converge to the slow manifold
\begin{equation}
    n=\frac{\alpha+s(b-c)p_0}{\lambda}\approx \frac{\alpha}{\lambda}=M,
    \label{eq:slow manifold}
\end{equation}
while $p$ does not change from $p_0$ (see example in Supplementary Fig.~S3B). After the population size reaches carrying capacity, trajectories then move along the slow manifold until one type or the other fixes ($p=0$ or $p=1$). We focus on the dynamics on the slow manifold, which simplifies the analysis to a one-dimensional system \cite{Parsons2017}. 

\begin{figure}[h!]
    \centering
    \includegraphics[scale=0.65]{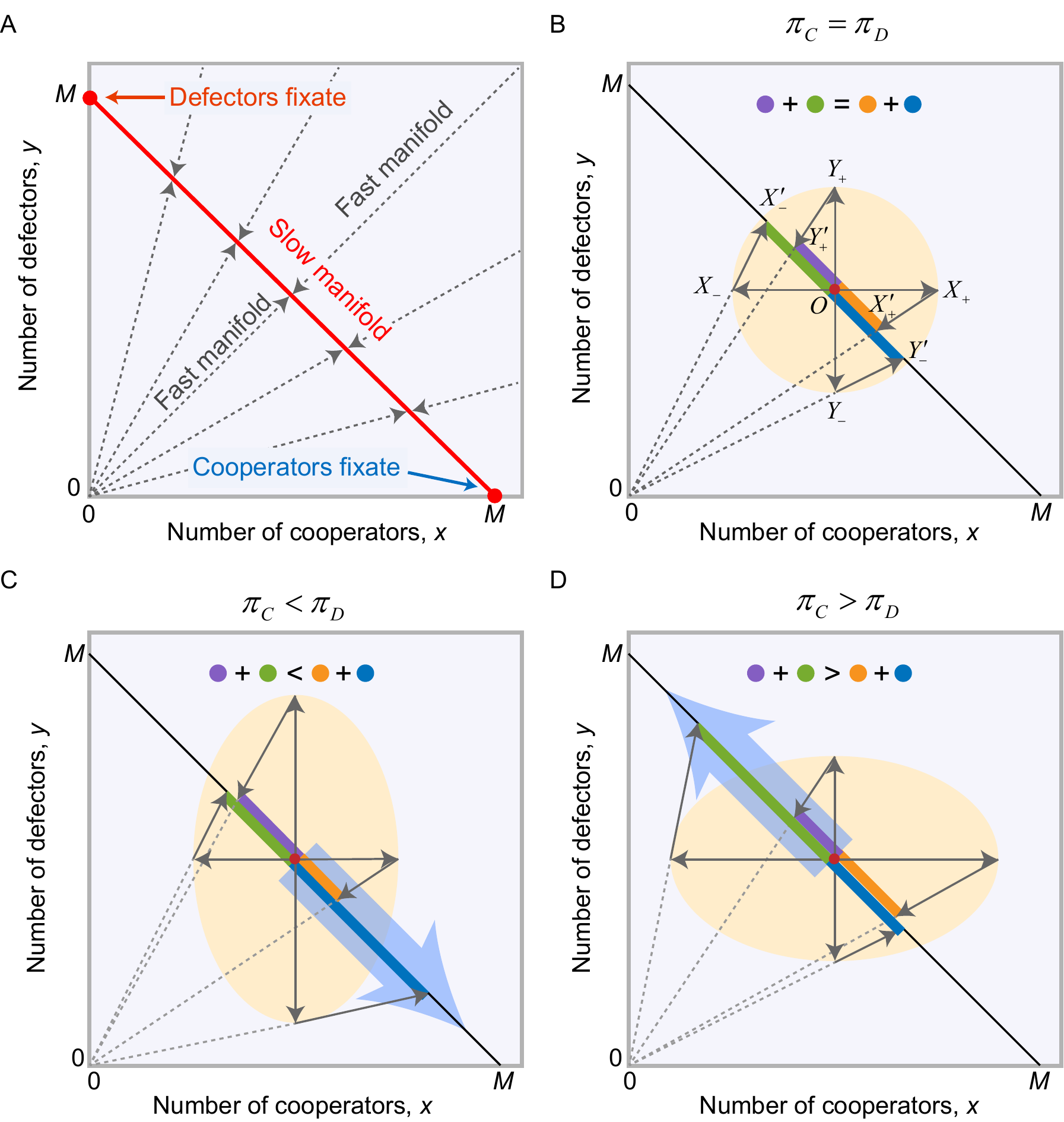}
    \caption{\textbf{How demographic stochasticity can favor cooperation or defection.}  
    (A) The system features a separation of timescales, where the total number of individuals $n=x+y$ changes much faster than the fraction of cooperators $p=x/(x+y)$.
    Starting from $x_0$ and $y_0$ cooperators and defectors, trajectories rapidly converge to the slow manifold ($x+y=M$) along the fast manifold $x/y=x_0/y_0$.
    (B, C, D) Stochastic fluctuations away from the slow manifold, followed by rapid return, can induce an advective force on the frequency of cooperators.
    For simplicity we consider constant payoffs, where $\pi_C$ and $\pi_D$ are independent of the number of cooperators and defectors. The ellipses illustrate the variance-covariance structure  of two-dimensional Gaussian fluctuations around the slow manifold from a given point $x=M/2$ and $y=M/2$ (red point $O$). 
    (B) When $\pi_C=\pi_D$, fluctuations from point $O$ are isotropic, shown as a circle.
    We consider four representative fluctuations from point $O$, $X_{-},X_{+},Y_{-},Y_{+}$, and the following points of return $X_{-}',X_{+}',Y_{-}',Y_{+}'$ to the slow manifold.
    For isotropic fluctuations there is no expected change in $p$ after return to the slow manifold.
    (C) For $\pi_C<\pi_D$, the Gaussian fluctuations are an-isotropic, shown as an ellipse, with larger fluctuations in the number of defectors. This asymmetry leads to an expected increase in cooperator frequency $p$ after return to the slow manifold, as indicated by the blue arrow.
    (D) For $\pi_C>\pi_D$, the larger fluctuation occurs in the number of cooperators, which leads to an expected decrease in cooperator frequency after return to the slow manifold. These effects of an-isotropic noise are similar to those discussed by \cite{Constable2016}, but they arise here even when both types have the same baseline birth rate and the same carrying capacity, under weak selection.}
    \label{fig4}
\end{figure}

In the co-ordinate system $(x,y)$, the slow manifold is defined $x+y=M$, and the fast manifolds are lines connecting the origin to points on the slow manifold (see Fig.~\ref{fig4}A). 
Given any initial conditions, the trajectory will rapidly approach the slow manifold along one of these lines, and then subsequently move within the slow manifold. However, unlike the case of a strictly constant population size, the system with demographic stochasticity does not lie precisely on the slow manifold at all times. Small fluctuations take the system off the slow manifold briefly, and then the system rapidly  returns to the slow manifold. Critically, the position where the system returns to the slow manifold, after a fluctuation, is not necessarily the same as where it started. In fact, there can be a systematic deviation in the position on the slow manifold that arises from stochastic fluctuations and rapid returns -- which produces an advective force on the frequency $p$ along the slow manifold (see Fig.~\ref{fig4}B, C, D). It is this systematic deviation, caused by demographic stochasticity, that introduces a force favoring cooperation.

In particular, fluctuations from $x+y=M$ follow a two-dimensional Gaussian distribution with variance $x(\delta_1B+\delta_2s\pi_C+D+\lambda n)$ in the $x$-direction and variance $y(\delta_1B+\delta_2 s\pi_D+D+\lambda n)$ in the $y$-direction. In Fig.~\ref{fig4}, we illustrate the fluctuation starting from state $x=y=M/2$ (see Supplementary Information Section S2.2 for the analysis of any other states).
When $\pi_C=\pi_D$, the Gaussian fluctuation is isotropic, and so a fluctuation followed by return along a fast-manifold line produces no expected change in the resulting position on the slow manifold (see Fig.~\ref{fig4}B). However, whenever $\pi_C \ne \pi_D$, the two-dimensional Gaussian fluctuation has an ellipsoid shape, and fluctuation followed by rapid return produces an expected change in the frequency of cooperators, $p$, along the slow manifold. In particular, when $\pi_C < \pi_D$, the expected change due to demographic stochastic favors cooperators, whereas if $\pi_C > \pi_D$ the expected change favors defectors (Fig.~\ref{fig4}C,D). In general, we can analytically calculate the adjective force along the slow manifold that arises from these stochastic fluctuations and rapid returns (Section S2.2 in Supplementary Information).

For the donation game we have studied so far, cooperators always have a lower payoff than defectors regardless of the population state. And so the advective force arising from demographic stochasticity always favors cooperation, regardless of $p$. If this force is large enough relative to the deterministic force favoring defectors, then it can produce a net advantage for cooperators. For other types of pairwise games, however, the direction of deterministic selection  ($\pi_C$ vs $\pi_D$) may depends on the current frequency $p$ in the population, and so the noise-induced advection may change sign along the slow manifold, producing complicated effects on long-term dynamics. We investigate these effects of demographic noise on evolutionary dynamics for general two-player games in the next section.

\subsection{General evolutionary game dynamics with demographic stochasticity}

For an arbitrary two-player game that gives rise to payoffs, the two-dimensional system can be simplified to a one-dimensional system by separation of timescales, provided selection is weak enough, $s \ll \alpha$.  Suppose the game has the following payoff structure:
\begin{equation}
    \begin{array}{cc}
       & \begin{array}{cc}
          {\rm C} & {\rm D} 
      \end{array} \\ \begin{array}{c}
           {\rm C}\\{\rm D} 
      \end{array}
     & \left(\begin{array}{cc}
          a&b \\
          c&d 
     \end{array}\right).
\end{array}
\end{equation}
Players have two strategies, which we still generically call cooperation (C) or defection (D). When two cooperators interact, both of them receive payoff $a$. When a cooperator interacts with a defector, the cooperator receives $b$ and the defector $c$. Mutual defection brings payoff $d$ to both players. The average payoff for a cooperator or defector in a population are respectively
\begin{equation}
\begin{split}
    \pi_C&=\frac{xa+yb}{x+y}, \\
    \pi_D&=\frac{xc+yd}{x+y}.
\end{split}
\end{equation}

Similar to Section \ref{section3.1}, we can describe the system by a stochastic differential equation:
\begin{subequations}
\begin{align}
          {\rm d}p=&sp(1-p)\left(1-\frac{\delta_2}{n}\right)(\pi_{\rm C}-\pi_{\rm D}){\rm d}t+\frac{1-p}{n} \sqrt{x(\delta_1 B+D+\lambda n)}{\rm d}W^{(1)}_t \notag \\
          &-\frac{p}{n}\sqrt{y(\delta_1 B +D+\lambda n)}{\rm d}W^{(2)}_t, \label{eq:transformed_general_a}\\
          {\rm d}n=&[n\alpha+s(p\pi_{\rm C}+(1-p)\pi_{\rm D})pn-\lambda n^2]{\rm d}t+ \sqrt{x(\delta_1 B+D+\lambda n)}{\rm d}W^{(1)}_t\notag \\
          &+\sqrt{y(\delta_1 B +D+\lambda n)}{\rm d}W^{(2)}_t. \label{eq:transformed_general_b}
\end{align}
\label{eq:transformed_general}
\end{subequations}
Since the population size quickly equilibrates to the carrying capacity  $M\approx\alpha/\lambda$, we substitute $n=M$ into Eq.~\ref{eq:transformed_general_a} which yields a one-dimensional equation for the evolution of $p$ along the slow manifold:
\begin{subequations}
\begin{align}
    {\rm d}p=&sp(1-p)\left[\left(1-\frac{\delta_2}{M}\right)\left(b-d+(a-b-c+d)p\right) \right]{\rm d}t \notag \\
    &+\sqrt{\frac{(\delta_1+1) B p(1-p)}{M}}\left(\sqrt{1-p}{\rm d}W^{(1)}_t-\sqrt{p}{\rm d}W^{(2)}_t\right).
\end{align}
\label{eq:general one dimensional}
\end{subequations}

In the case of deterministic births and deaths ($\delta_1=\delta_2=0$ and neglecting variance in the death process), this equation simplifies to the classic replicator equation \cite{Schuster1983,Nowak2006a}. For general games there may be interior equilibrium points, and so the fixation probability is no longer a good measure to describe long-term evolutionary outcomes. Instead, we analyze the dynamics from two perspectives. One is from the perspective of the deterministic behavior on the slow manifold,
which neglects stochasticity altogether in Eq.~\ref{eq:general one dimensional} and studies the equilibria of the resulting ordinary differential equation. The other, more nuanced perspective accounts for stochasticity. Since $p=0$ and $p=1$ are the only absorbing states, any trajectory will finally reach one of these states and then become invariant. However, we can impose a reflecting condition on the boundary, which is equivalent to assuming that, when the number of one phenotype reaches zero, a new mutant of this phenotype arises instantly. The resulting evolutionary process of $p$ becomes an ergodic Markov process which has a unique stationary distribution $v^*(p)$. A frequency $p$ with greater probability density means that trajectories spend more time there. Derivation of the stationary distribution $v^*(p)$ under reflecting boundaries is given in Section S4.1 of Supplementary Information. 

When we ignore the stochastic terms, then Eq.~\ref{eq:general one dimensional} is an ODE with the same equilibrium points and stabilities as the classic replicator equation, provided $\delta_2<M$. Whereas if $\delta_2>M$, then the equilibrium points are the same as the classic replicator equation, but the stabilities are reversed: equilibrium points that are classically unstable become stable, and conversely. And so the value of $\delta_2$, which determines the payoff-component of offspring variance, can reverse the evolutionary outcome, even from a deterministic perspective.

\begin{figure}[h!]
    \centering
    \includegraphics[width=\textwidth]{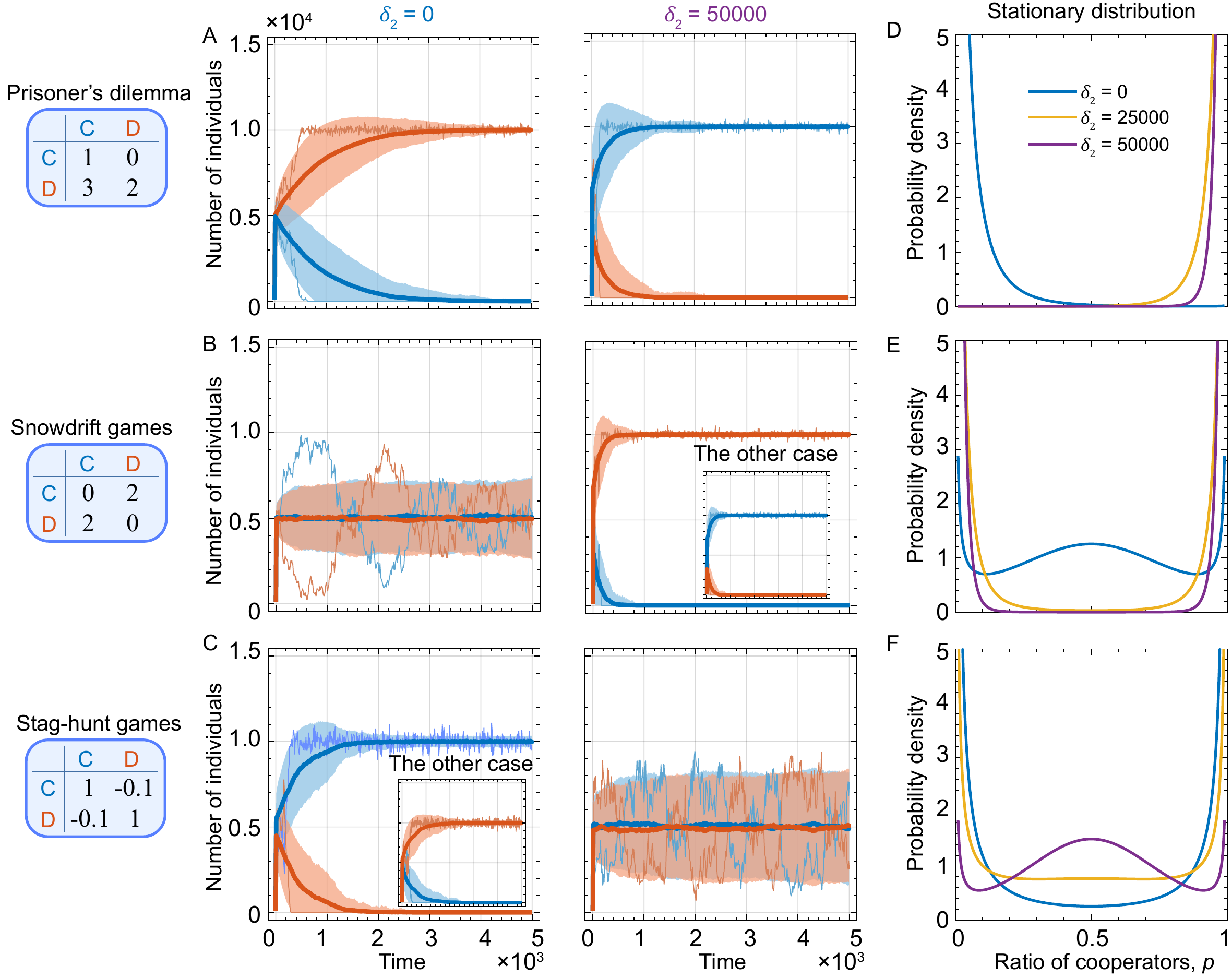}
    \caption{\textbf{General evolutionary game dynamics with demographic stochasticity.} 
    We consider three types of representative games, such as prisoner's dilemma (A, D), snowdrift game (B, E), and stag-hunt games (C, F).
    In the prisoner's dilemma games, when demographic stochasticity is absent or does not meet $\delta_2>M$, defectors dominate the population (see trajectories sampled in A, left part).
    While the evolutionary direction can be reversed for $\delta_2>M$, where cooperation becomes the dominant strategy (see trajectories sampled in A, right part).
    Shown in (D) is the stationary distribution of cooperators for $\delta_2=0$, $\delta_2=25000$, and  $\delta_2=50000$.
    Analogously, in the snowdrift game, the demographic stochasticity with $\delta_2>M$ changes the equilibrium from the coexistence of two strategies (B, left part) to the bi-stability (B, right part), which suggests the transformation of a snowdrift game to a stag-hunt game.
    We also find that with demographic stochasticity, the evolution in the stag-hunt games proceed ``as if" the population are playing snowdrift games. 
    Parameters: $B=2$, $D=1$, $s=10^{-3}$, $\delta_1=2.5$, $\lambda=10^{-4}$, $x_0=y_0=100$. }
    \label{fig5}
\end{figure}

More generally, we can classify three different deterministic scenarios based on the payoff matrix of the two-player, two-action game. For dominance games (Fig.~\ref{fig5}A), one strategy is always dominant. Here, without loss of generality, we assume defection dominates cooperation ($a<c$ and $b<d$, e.g.,~a prisoner's dilemma). If $\delta_2<M$, then all trajectories will converge to the full-defector state ($p=0$ stable and $p=1$ unstable). However, if $\delta_2>M$, cooperation becomes the dominant strategy and all trajectories converge to full-cooperator state ($p=1$ stable and $p=0$ unstable). For coexistence games ($a<c$ and $b>d$, e.g.,~a snowdrift game), the best response is to choose the opposite strategy of the opponent (Fig.~\ref{fig5}B). If $\delta_2<M$, there is only one stable equilibrium, $p^*=(d-b)/(a-b-c+d)$. All trajectories will converge to $p^*$ and therefore cooperators and defectors stably coexist. If $\delta_2>M$, $p^*$ becomes unstable and $p=0$ and $p=1$ are each stable. Thus, all trajectories converge to either the full-cooperator or the full-defector state, similar to the outcome of a classic coordination game. For coordination games (Fig.~\ref{fig5}C), the best response is to choose the same strategy as the opponent ($a>c$ and $d>b$, e.g., a stag-hunt game). In this case, $\delta_2<M$ leads to an unstable internal equilibrium $p^*$ with stable boundaries ($p=0$ and $p=1$). But for $\delta_2>M$, $p^*$ becomes stable while $p=0$ and $p=1$ are unstable. Most trajectories fluctuate around $p^*$ for a long time, showing similar behavior as a classic coexistence game. In summary, in a population with sufficiently large offspring variance ($\delta_2>M$), the outcome of each type of game has the dynamical properties classically associated with the opposite type of game in a deterministic setting. In other words, demographic stochasticity effectively transforms the payoff structure of a game in the following way

\begin{equation}
    \begin{pmatrix}
    a&b\\c&d
    \end{pmatrix} \Rightarrow \begin{pmatrix}
    -a&-b\\-c&-d
    \end{pmatrix}.
\end{equation}

We can also characterize general two-player games in term of the stationary frequency distribution of strategies, with reflecting boundaries. This description accounts for more details in the stochastic dynamics, and it reveals a similar, transformative effect of large offspring variance. 
If $\delta_2$ is sufficiently large, namely $\delta_2>M$, then modes of the stationary distribution can be moved from one boundary to the other boundary (dominance games, Fig.~\ref{fig5}D), from the interior to the two boundaries (coexistence games, Fig.~\ref{fig5}E), or from the two boundaries to the interior (coordination games, Fig.~\ref{fig5}F). 
These results reflect our ODE-based analysis above, and they show that sufficient offspring variance can reverse the evolutionary dynamics in an interacting population. These dramatic effects extend to games with more than two actions, such as rock-paper-scissors (Supplementary Fig.~S4).

These two analytical perspectives  underscore that large offspring variance can reshape the payoff structure of a game, producing dynamics classically seen in an entirely different game type.  So far, we have focused on the scaling factor $\delta_2$, which governs how offspring variance grows with payoff, as opposed to $\delta_1$, which governs the baseline offspring variance. The value of $\delta_1$ can also profoundly influence evolutionary outcomes, although this cannot be seen from a deterministic perspective alone because $\delta_1$ has no effect on stabilities of equilibria. Analysis of the stationary distribution shows that a large baseline variance ($\delta_1B$) can transform any game into a coordination game (see Section S4.1 in Supplementary Information). An example of this result is shown in Fig.~\ref{fig5}F, where even though $\delta_2=25,000$ exceeds the carrying capacity, the stationary distribution is not unimodal around intermediate frequency. This is because the effect of  $\delta_2$ here is offset by the effect of $\delta_1$.
These results show that demographic noise, especially when offspring variance is high, can qualitatively change the evolutionary outcomes compared to  predictions of traditional analysis by replicator equations for fixed or infinite population size \cite{Hofbauer1998}.

\section{Discussion}

The question of how cooperation can be maintained is a longstanding and active area of research, spanning multiple disciplines. A large literature has produced compelling explanations for cooperation, but these typically rely on some form of population structure or repeated interactions. Here, we find that even in a well-mixed population with one-shot interactions, natural stochasticity in the total population size alone can favor cooperation that would otherwise be suppressed. For other types of social interactions, as well, demographic stochasticity can reverse the direction of evolutionary trajectories and produce behavioral outcomes that contravene classical expectations.

It is intuitively easier to invade a noisy population than a stable population. And so natural selection near carrying capacity prefers types not only with higher fecundity (greater mean offspring number), but also with lower reproductive noise (smaller offspring variance) \cite{Parsons2007,Parsons2007a}. The reversal in the direction of selection in a stochastic population reflects this basic trade-off between offspring mean and offspring variance. A larger payoff produces higher fecundity but also greater noise in the reproduction process. Whether it is the mean or the variance in offspring number that dominates the course of evolution is determined by their relative importance, which is governed by $\delta_2$ in our model.
Classical models of populations with constant (or infinite) size neglect the effects of offspring variance altogether; but more realistic models, we have seen, permit regimes where offering variance is more important than fecundity.

Although demographic noise has been studied extensively in population models, the underlying mechanism for our results is qualitatively different from those explored in prior studies. Most research on demographic noise has been restricted constant fitness for competing types \cite{Parsons2007,Parsons2007a,Parsons2010,McKane2005,Butler2009,Hallatschek2007,Stollmeier2018,Wienand2017,Taitelbaum2020,Chotibut2017}, which does not provide a model of social interactions. However, Constable et al.~ analyzed a frequency-dependent fitness model, and they also found that demographic noise can reverse the direction of selection \cite{Constable2016}. Their model is based on the production and consumption of a public good. One phenotype produces the public good, at a cost that reduces its baseline birth rate, while the other phenotype does not produce the public good. 
They analyze the case when ``cooperators" (who produce the public good) have a larger intrinsic carrying capacity than non-producers, and the larger carrying capacity then yields an evolutionary advantage by making producers more robust against invasion. 
This mechanism is thus a stochastic form of $r$ versus $K$ selection \cite{pianka1970r}, and it occurs when births and deaths follow Poisson processes. By contrast, in our model, the evolutionary advantage of cooperators arises even though both types have the same baseline birth rate and the same carrying capacity; 

and it arises only when the birth process related to payoff is sufficiently over-dispersed. 
This mechanism is thus fundamentally different from a trade-off between baseline birth rate and carrying capacity of competing types in a Poisson model \cite{Constable2016,Houchmandzadeh2012,Houchmandzadeh2015}, and it is more closely related to phenomena in population models with heavy-tailed offspring distributions \cite{schweinsberg2000necessary,Eldon2006,Sargsyan2008,der2012dynamics}.

Aside from promoting cooperation in the prisoner's dilemma, demographic stochasticity also transforms outcomes in other forms of social interaction. Stochasticity can convert a snowdrift game into a stage-hunt game, for example, so that the stable co-existence expected in a deterministic or Poisson setting is transformed into bi-stability. Here, again, the underlying mechanism that reverses the evolutionary outcome is over-dispersion in the offspring contribution related to payoff, even when both types have the same baseline birth rate and carrying capacity.

All of our analyses have assumed a fast-growing population ($\alpha \gg s$), which rapidly reaches carrying capacity before any change in the relative frequencies of competing types. The dynamics of competition may be more complicated in a stochastic, slow-growing population, because their analysis cannot be reduced to a one-dimensional slow manifold. In this regime, fixation will take place before reaching carrying capacity. We can nonetheless derive approximations for the fixation probability in this regime as well (Section S4.2 in Supplementary Information), and, in the case of the donation game, we find that cooperation will be favored by selection provided $\delta_2$ exceeds the initial population size, $\delta_2>n_0$. This condition is typically easier to satisfy than Eq.~\ref{eq:condition}, and it is confirmed by both numerical simulations and Monte Carlo simulations of the compound Poisson process (Supplementary Fig.~S5 and Fig.~S6). After cooperators or defectors fix, in this regime of a slow-growing population, the population will tend to grow logistically to its carrying capacity; but in this case the carrying capacity is larger for cooperators (Supplementary Fig.~S7), which provides an additional evolutionary advantage and greater chance of long-term persistence (Supplementary Fig.~S8).

Our results highlight the strong impact of stochasticity on evolutionary outcomes in populations. The demographic stochasticity we have studied arises from intrinsic properties of birth and death processes, which have size of order $O(\sqrt{n})$. As the population size grows towards infinity this form of stochasticity has little influence on evolutionary dynamics, which is consistent with the recent finding that migration in finite, group-structured populations can favor cooperators provided the population size is not too large \cite{Braga2022}. Aside from intrinsic stochasticity during reproduction, real populations may also be subject to external noise, arising from exogenous variation in environmental conditions. Unlike demographic noise, exogenous noise  can be substantial even in population of arbitrary large size. Prior studies on environmental fluctuations, including fluctuations in selection intensity \cite{Assaf2013}, carrying capacity \cite{Wienand2017,Taitelbaum2020}, and payoff structure \cite{Stollmeier2018}, have analyzed their effects by imposing an external noise term onto an otherwise classical, deterministic and continuous system of equations.  The effects of exogenous noise on discrete stochastic systems remain less explored, and they are likely to differ qualitatively from stochastic perturbations of continuous systems  \cite{Durrett1994}. Coupling intrinsic demographic noise with external environmental noise may produce even more complicated effects, which remains a topic for future research.

The impact of stochasticity on strategic outcomes likely extends beyond the two-player/two-action games we focused on, to include many aspects of non-human and human social behavior. Even if behavioral spread is caused by biased imitation, there is nonetheless variance in number of individuals who imitate a type, as well as physical variation in population sizes of interacting social groups as individuals move between social settings. Empirical data has documented burstiness, a form of over-dispersion, in  social interactions \cite{Stehle2010,Goh2008}. Likewise, in the context of behavior during an epidemic, there is evidence of super-spreading individuals that cause over-dispersion in infectiousness \cite{Tkachenko2021,Kirkegaard2021}, which may influence frequency-dependent competition among co-circulating variants. Extending our model and analysis to these settings remains an open topic for future research.

\clearpage
\bibliographystyle{unsrtnat}
\bibliography{references}



\section*{Supplementary material}

# The arrow of evolution when the offspring variance is large
# Supplementary Information

Guocheng Wang[1], Qi Su[2,3,4], Long Wang[1,4], and Joshua B. Plotkin[2,3,5]

[1]Center for Systems and Control, College of Engineering, Peking University, Beijing 100871, China
[2]Center for Mathematical Biology, University of Pennsylvania, Philadelphia, PA 19104, USA
[3]Department of Mathematics, University of Pennsylvania, Philadelphia, PA 19104, USA
[4]Center for Multi-Agent Research, Institute for Artificial Intelligence, Peking University, Beijing 100871, China
[5]Department of Biology, University of Pennsylvania, Philadelphia, PA 19104, USA

September 6, 2022

# Contents

# S1 Basic assumptions and population model

We use a continuous-time, discrete state Markov process to describe births and deaths in a population of replicating individuals. We study a family of population models by making the following axiomatic assumptions about the birth process:

**Assumption S1.1.** *Let the stochastic process $X_t^{(i)}$ denote the number of offspring that individual i produces in time $0 \to t$. We assume that the stochastic process satisfies the following four properties:*
*(1) $X_0^{(i)} = 0$;*
*(2) Independent increment: $X_{t+\tau}^{(i)} - X_t^{(i)}$ is independent of $X_t^{(i)}$;*
*(3) Stationary increment: $X_{t+\tau}^{(i)} - X_t^{(i)}$ and $X_\tau^{(i)}$ have the same distribution;*
*(4) Non-decreasing: $X_{t+\tau}^{(i)} - X_t^{(i)} \geq 0$.*

Given these properties, the mean and variance of $X_t^{(i)}$ are both linear functions of $t$, and the offspring distribution (the number of offspring within unit of time) is homogeneous and independent of time. We use $\xi_i$ to denote the offspring distribution. To describe the distribution of $\xi_i$ we make the following axiomatic assumptions about its expectation and variance:

**Assumption S1.2.** *The mean and variance of $\xi_i$ are given by*
*(1) $\mathbb{E}(\xi_i) = B + s\pi_i$;*
*(2) $\text{Var}(\xi_i) = \delta_1 B + \delta_2 s\pi_i$.*

Here, the parameter $B$ denotes the baseline of birth rate, $\pi_i$ denotes individual $i$'s payoff (obtained from pairwise social interactions in the population), and $s$ denotes the selection intensity. A larger value of $s$ means that payoffs have a stronger effect on individual's birth process. In this paper, we adopt the common assumption that the selection intensity is weak ($s \ll 1$).

The expected number of offspring in one unit of time, $\mathbb{E}(\xi_i)$, is called the fitness of individual $i$. In the regime of weak selection we can expand $\mathbb{E}(\xi_i)$ as a Taylor series and truncate it to first order in $\pi_i$. Thus, Assumption S1.2 includes a large range of different birth models. Note that in a classical discrete-time model of reproduction, such as the Moran process, fitness is often assumed to be an exponential function of payoff, i.e, $\exp(B + s\pi_i)$. This is consistent with our definition of fitness $\mathbb{E}(\xi_i) = B + s\pi_i$ in continuous time.

The space of models defined by the axioms above includes models in which the birth process is not Poisson, but is rather over-dispersed. There are many ways to produce over-dispersion, where the mean and variance are correlated in different ways [1, 2], such as quasi-Poisson model (variance is proportional to mean), mixed-effects Poisson model and negative binomial model (variance is a quadratic function of mean). Here, we stipulate only a general form of correlation between variance and mean, which covers a large range birth different processes: $\text{Var}(\xi_i) = f(\mathbb{E}(\xi_i)) = f(B + s\pi_i)$. In the regime of weak selection $s \ll 1$,

we can expand the variance to first order in $s$, which yields (2) in Assumption S1.2. The parameter $\delta_1$ measures how the variance component in the offspring distribution scales with the baseline birth rate, which is the same for all individuals. Whereas the parameter $\delta_2$ measure how the variance component of the offspring distribution scales with the current payoff of individual $i$, which may vary across individuals. The case $\delta_1 = \delta_2 = 1$ simplifies to a classic Poisson birth process. Whereas if $\delta_1 > 1$, the birth processes is over-dispersed (because we work in the regime $s \ll 1$).

Suppose there are $N$ individuals of a given phenotype. Thus these $N$ individuals have the same but independent birth rates. The overall birth rate of this phenotype is then

$$\sum_{i=1}^{N} \xi_i, \tag{S1.1}$$

which converges to a normal distribution when $N$ is large, by the central limit theorem. We approximate the Markov process by a diffusion equation, which depends only on the first two moments of $\xi_i$ and is expected to be a good approximation in large enough populations. In addition, we perform explicit simulations of processes, and compare their results to the predictions derived by the diffusion approximation, in Section S3.

We assume death events follow a Poisson process governed by two terms, for each phenotype. One term arises from a constant baseline death rate, denoted by $D$. The other term corresponds to deaths caused by competition among individuals for limited resources, which introduces a carrying capacity on the entire population. Specifically, the competition process is modelled by a reaction process

$$X + Y \to X, \tag{S1.2}$$

which takes place with constant rate $\lambda$ and $X$ and $Y$ are any two individuals. Thus, the death rate caused by competition is $\lambda$ times the population size.

Given these assumptions, the population is described by a continuous-time discrete state Markov process, which can be approximated by a diffusion equation. We start by using the traditional donation game as an example, to produce payoff expressions and derive the corresponding diffusion equation. Consider a population composed of $x$ cooperators and $y$ defectors. Each individual plays games with all other players (including self) and obtains an average payoff. Each cooperator provides a benefit $b$ to his opponent and pays a cost $c$, but a defector contributes nothing and pays no cost. The payoff matrix for a pairwise interaction is as follows:

$$\begin{array}{c} \\ C \\ D \end{array} \begin{pmatrix} \overset{C}{b-c} & \overset{D}{-c} \\ b & 0 \end{pmatrix}. \tag{S1.3}$$

Summing over all pairwise interactions, the total payoff for a cooperator and defector are

$$\pi_C = \frac{x}{x+y}b - c, \tag{S1.4a}$$

$$\pi_D = \frac{x}{x+y}b. \tag{S1.4b}$$

We assume that in a short time interval $\Delta t$, the number of births of cooperators is $\Delta x^+$, and the number of deaths of cooperators is $\Delta x^-$. Thus the increment of cooperators is $\Delta x = \Delta x^+ - \Delta x^-$. The same applies to defectors, $y$. Then, we obtain the following expressions for mean change in state of short time $\Delta t$

$$\mathbb{E}[\Delta x(t)|x(t), y(t)] = x(t)[B + s\pi_C - D - \lambda(x(t) + y(t))]\Delta t, \tag{S1.5a}$$

$$\mathbb{E}[\Delta y(t)|x(t), y(t)] = y(t)[B + s\pi_D - D - \lambda(x(t) + y(t))]\Delta t. \tag{S1.5b}$$

Since deaths follow a Poisson process, the variance of $\Delta x^-$ equals to its mean, namely $Dx\Delta t + \lambda x(x+y)\Delta t$. Moreover, $\Delta x^+$ is independent of $\Delta x^-$, which leads to the following expressions for the variance in state change of short time $\Delta t$

$$\begin{aligned}\mathbb{E}[\Delta x(t)^2|x(t), y(t)] &\approx \text{Var}[\Delta x(t)^+ - \Delta x(t)^-] = \text{Var}[\Delta x(t)^+] + \text{Var}[\Delta x(t)^-] \\ &= x(t)(\delta_1 B + \delta_2 s\pi_C)\Delta t + Dx(t)\Delta t + \lambda x(t)(x(t) + y(t))\Delta t + o(\Delta t),\end{aligned} \tag{S1.6a}$$

$$\begin{aligned}\mathbb{E}[\Delta y(t)^2|x(t), y(t)] &\approx \text{Var}[\Delta y(t)^+ - \Delta y(t)^-] = \text{Var}[\Delta y(t)^+] + \text{Var}[\Delta y(t)^-] \\ &= y(t)(\delta_1 B + \delta_2 s\pi_D)\Delta t + Dy(t)\Delta t + \lambda y(t)(x(t) + y(t))\Delta t + o(\Delta t).\end{aligned} \tag{S1.6b}$$

Thus, we obtain

$$\lim_{\Delta t \to 0} \frac{\mathbb{E}[\Delta x(t)|x(t)=x, y(t)=y]}{\Delta t} = x[B + s\pi_C - D - \lambda(x+y)], \tag{S1.7a}$$

$$\lim_{\Delta t \to 0} \frac{\mathbb{E}[\Delta y(t)|x(t)=x, y(t)=y]}{\Delta t} = y[B + s\pi_D - D - \lambda(x+y)], \tag{S1.7b}$$

$$\lim_{\Delta t \to 0} \frac{\mathbb{E}[\Delta x(t)^2|x(t)=x, y(t)=y]}{\Delta t} = (\delta_1 B + \delta_2 s\pi_C)x + Dx + \lambda x(x+y), \tag{S1.7c}$$

$$\lim_{\Delta t \to 0} \frac{\mathbb{E}[\Delta y(t)^2|x(t)=x, y(t)=y]}{\Delta t} = (\delta_1 B + \delta_2 s\pi_D)y + Dy + \lambda y(x+y), \tag{S1.7d}$$

$$\lim_{\Delta t \to 0} \frac{\mathbb{E}[\Delta x(t)\Delta y(t)|x(t)=x, y(t)=y]}{\Delta t} = 0. \tag{S1.7e}$$

In what follows, for simplicity we use $\alpha$ to denote $B - D$, which represents the baseline net growth rate of the population, independent of payoff effects. Then we can then use the following diffusion equation to approximate the Markov process[3]:

$$dx = x[\alpha + s\pi_C - \lambda(x+y)]dt + \sqrt{(\delta_1 B + \delta_2 s\pi_C)x + Dx + \lambda x(x+y)}dW_t^{(1)}, \tag{S1.8a}$$

$$dy = y[\alpha + s\pi_D - \lambda(x+y)]dt + \sqrt{(\delta_1 B + \delta_2 s\pi_D)y + Dy + \lambda y(x+y)}dW_t^{(2)}, \tag{S1.8b}$$

where $W_t^{(1)}$ and $W_t^{(2)}$ are two independent Wiener processes, and the diffusion terms are interpreted in the sense of Itô.

## S2 Methods

### S2.1 Fixation probability

To investigate the dynamics of population size and cooperator frequency, we introduce a coordinates transformation, setting $p = x/(x+y)$ and $n = x+y$. Applying Itô's lemma

$$\mathrm{d}p = \frac{\partial p}{\partial x}\mathrm{d}x + \frac{\partial p}{\partial y}\mathrm{d}y + \frac{1}{2}\frac{\partial^2 p}{\partial x^2}(\mathrm{d}x)^2 + \frac{1}{2}\frac{\partial^2 p}{\partial y^2}(\mathrm{d}y)^2, \tag{S2.1a}$$

$$\mathrm{d}n = \frac{\partial n}{\partial x}\mathrm{d}x + \frac{\partial n}{\partial y}\mathrm{d}y + \frac{1}{2}\frac{\partial^2 n}{\partial x^2}(\mathrm{d}x)^2 + \frac{1}{2}\frac{\partial^2 n}{\partial y^2}(\mathrm{d}y)^2, \tag{S2.1b}$$

to Eq. S1.8 we obtain

$$\begin{aligned}\mathrm{d}p =&\, scp(1-p)\left(\frac{\delta_2}{n} - 1\right)\mathrm{d}t + \frac{1-p}{n}\sqrt{x(\delta_1 B + D + \lambda n)}\mathrm{d}W_t^{(1)} \\ &- \frac{p}{n}\sqrt{y(\delta_1 B + D + \lambda n)}\mathrm{d}W_t^{(2)},\end{aligned} \tag{S2.2a}$$

$$\begin{aligned}\mathrm{d}n =&\, [n\alpha + s(b-c)pn - \lambda n^2]\mathrm{d}t + \sqrt{x(\delta_1 B + D + \lambda n)}\mathrm{d}W_t^{(1)} \\ &+ \sqrt{y(\delta_1 B + D + \lambda n)}\mathrm{d}W_t^{(2)}.\end{aligned} \tag{S2.2b}$$

Here, we have ignored terms of order $O(s)$ in the diffusion term.

We work in the regime of weak selection, $s \ll 1$. Here the population growth rate is mainly determined by the baseline birth and death rates, $\alpha = B - D$. If $\alpha > 0$, the population will grow logistically until it reaches the carrying capacity. If $\alpha < 0$, the population will eventually perish. The case of $\alpha = 0$ is a critical case, where the growth rate is totally determined by payoffs. In this paper, we only discuss the case of $\alpha > 0$.

In this section, we focus first on a fast-growing population ($\alpha \gg s$). In this setting, the system features a time-scale separation [4]. For the corresponding ordinary differential equation (i.e. omitting the diffusion term in Eq. S2.2), since $\alpha \gg s$, $n$ equilibrates much more quickly than $p$. Therefore, given an arbitrary initial configuration $p_0$ and $n_0$, before $p$ changes at all the dynamical system rapidly converges to the slow manifold defined by $\mathrm{d}n = 0$, or more explicitly,

$$n = \frac{\alpha + s(b-c)p}{\lambda} \approx \frac{\alpha}{\lambda}. \tag{S2.3}$$

The slow manifold is in fact the carrying capacity of the environment, denoted by $M$. It reflects the maximum number of individuals that the environment can sustain. For the stochastic system, the system still shows the features of fast-slow dynamics (Fig. S3). The population size $n$ will quickly converges to $M$, and then fluctuate around it.

Since the population reaches the carrying capacity rapidly, and the population size then fluctuate around $M$ in subsequent evolution, we can use $M$ to replace $n$ in Eq. S2.2a. Remembering $D + \lambda M = B$, we obtain a one-dimensional diffusion system describing the evolution of cooperator frequency on the slow manifold:

$$\mathrm{d}p = scp(1-p)\left(\frac{\delta_2}{M} - 1\right)\mathrm{d}t + \sqrt{\frac{(\delta_1 + 1)Bp(1-p)}{M}}\left(\sqrt{1-p}\mathrm{d}W_t^{(1)} - \sqrt{p}\mathrm{d}W_t^{(2)}\right). \tag{S2.4}$$

5

The solution of Eq. S2.4 is a Markov process with the infinitesimal generator given by

$$\mathcal{L}f = scp(1-p)\left(\frac{\delta_2}{M}-1\right)\frac{\partial f}{\partial p} + \frac{(\delta_1+1)Bp(1-p)}{2M}\frac{\partial^2 f}{\partial p^2}. \tag{S2.5}$$

For Eq. S1.8, $x = 0$ and $y = 0$ are two absorbing states. Given the initial number of cooperators $x_0$ and defectors $y_0$, for some time $t$, if $x(t) > 0$ and $y(t) = 0$ (i.e. $p = 1$) is satisfied, we say cooperators have fixed [5]. The probability that cooperators will fix is called fixation probability, denoted $\rho$. Given the initial frequency of cooperators $p_0$ ($p_0 = x_0/(x_0+y_0)$), the fixation probability of cooperators $\rho(p_0)$ is the solution of

$$\begin{cases} \mathcal{L}\rho = 0, \\ \rho(1) = 1, \\ \rho(0) = 0. \end{cases} \tag{S2.6}$$

The solution of this equation can be calculated explicitly [3], which is

$$\rho(p_0) = \frac{\int_0^{p_0} S(x)\mathrm{d}x}{\int_0^1 S(x)\mathrm{d}x}, \tag{S2.7}$$

where

$$S(x) = \exp\left(\int -\frac{2E(p)}{D(p)}\mathrm{d}p\right), \tag{S2.8a}$$

$$E(p) = scp(1-p)\left(\frac{\delta_2}{M}-1\right), \tag{S2.8b}$$

$$D(p) = \frac{(\delta_1+1)Bp(1-p)}{M}. \tag{S2.8c}$$

Using this method, we obtain the fixation probability:

$$\rho(p_0) = \frac{\exp\left[\frac{2sc}{(\delta_1+1)B}(M-\delta_2)p_0\right]-1}{\exp\left[\frac{2sc}{(\delta_1+1)B}(M-\delta_2)\right]-1} \approx p_0 - \frac{sc}{(\delta_1+1)B}(M-\delta_2)p_0(1-p_0). \tag{S2.9}$$

If $s \to 0$, $\rho(p_0) \to p_0$, which means that fixation probability in the absense of selection is equal to the initial frequency of cooperators. Furthermore, if $2scM/(\delta_1B+B) - 2\delta_2sc/(\delta_1B+B) < 0$, i.e.

$$M < \delta_2, \tag{S2.10}$$

the fixation probability will exceed that of neutral drift, i.e. $\rho(p_0) > p_0$, which means that cooperation is favored by natural selection.

### S2.2 Intuition for the effects of demographic stochasticity

In Fig. 4 of the main text, we have illustrated that fluctuations taking the trajectories off the slow manifold quickly return to the slow manifold, travelling along one of the the fast manifolds. However, the position where the trajectories return to the slow manifold

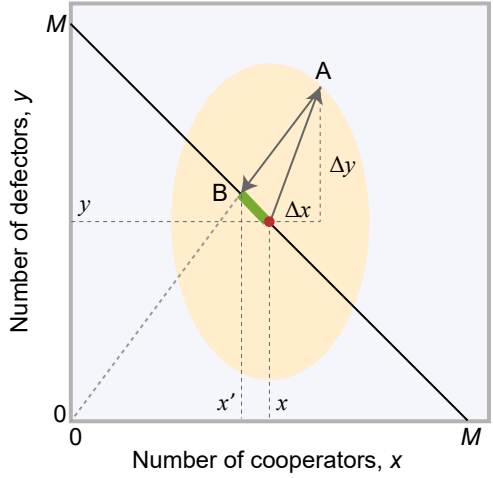

**Figure S1.** The advective force along the slow manifold induced by demographic noise.

is expected to have a deviation from the original starting point, which yields an effective advective force on the evolution of $p$ along the slow manifold.. Here, we provide a more accurate calculation for this drift.

Suppose the initial number of cooperators and defectors are $x$ and $y$, lying on the slow manifold $x + y = M$. As shown in Fig. S1, due to the noise, the trajectory may fluctuate to point $A$ (with fluctuation $\Delta x$ in $x$-direction and $\Delta y$ in $y$-direction) within a short time $\Delta t$. Then, it will return to the slow manifold at point $B$ along the fast manifold, which is very fast so that the return time can be ignored. As a result there exists an effective advective force on the number of cooperators (i.e. $x' - x$) along the slow manifold. We now derive the expectation of $x' - x$.

By Fig. S1, we can obtain that

$$x' = (x+y)\frac{x + \Delta x}{x + y + \Delta x + \Delta y}. \tag{S2.11}$$

Then, we have

$$\begin{aligned}
\mathbb{E}[x' - x] &= \mathbb{E}\left[\frac{x + \Delta x}{1 + \frac{\Delta x + \Delta y}{x+y}} - x\right] \\
&\approx \mathbb{E}\left[(x + \Delta x)\left(1 - \frac{\Delta x + \Delta y}{x+y} + \frac{(\Delta x + \Delta y)^2}{(x+y)^2}\right) - x\right] \\
&= \mathbb{E}\left[-\frac{y}{(x+y)^2}\Delta x^2 + \frac{x}{(x+y)^2}\Delta y^2\right].
\end{aligned} \tag{S2.12a}$$

Here, $(\Delta x, \Delta y)$ obeys a two-dimensional Gaussian distribution. According to Eq. S1.7, we have

$$\mathbb{E}[\Delta x^2] \approx \text{Var}(\Delta x) = x(\delta_1 B + \delta_2 s\pi_\text{C} + D + \lambda(x+y))\Delta t, \tag{S2.13a}$$

$$\mathbb{E}[\Delta y^2] \approx \text{Var}(\Delta y) = y(\delta_1 B + \delta_2 s\pi_\text{D} + D + \lambda(x+y))\Delta t. \tag{S2.13b}$$

7

Substituting Eq. S2.13 into Eq. S2.12, we have

$$\mathbb{E}[x' - x] = \frac{xy}{(x+y)^2}\delta_2 s(\pi_D - \pi_C)\Delta t. \quad (S2.14)$$

Thus, when $\pi_C < \pi_D$, the advective force induced by noise tends to increase the number of cooperators. When $\pi_D < \pi_C$, the number of cooperators is expected to decreases, and thus defectors are expected to increase. This analysis verifies the conclusion in Fig. 4 in the main text.

## S3  Examples: some explicit birth-death processes

We have analyzed a large class of models by stipulating axiomatic forms for the mean and variance in the birth process, and studying it by approximation using a diffusion equation. Here, for concreteness sake, we specify some explicit discrete-state/continuous times processes that satisfy our axiomatic assumptions (Assumption S1.1 and Assumption S1.2) and we demonstrate that our analysis predicts their behavior.

### Poisson process

Most prior studies depict birth events often through a reaction equation

$$Y \xrightarrow{\theta} Y + Y, \quad (S3.1)$$

where $Y$ is an arbitrary individual and $\theta$ is the birth rate. This assumption produces a birth process that is Poisson with intensity $\theta$. Within a time interval $[t, t + \Delta t]$, the offspring number of the individual obeys a Poisson distribution with expectation $\theta_i \Delta t$ and variance $\theta_i \Delta t$. As usual, the variance is equal to the mean in this classic case.

In our more general family of models we consider the number of offspring $\xi_i$ produced by individual $i$ within one unit of time, which is assumed to satisfy

$$\mathbb{E}(\xi_i) = B + s\pi_i = \theta_i, \quad (S3.2a)$$

$$\mathrm{Var}(\xi_i) = \delta_1 B + \delta_2 s\pi_i = \theta_i. \quad (S3.2b)$$

Our family of models thus contains the classical Poisson birth process in the case case $\delta_1 = \delta_2 = 1$. In this case, the condition that cooperation is favored becomes to

$$M < 1, \quad (S3.3)$$

which can never be achieved since the carrying capacity can never be lower than 1. And so, under our family of models, a strictly Poisson birth process can never favor cooperation.

### Compound Poisson process

In Poisson process, there is only one offspring produced in each reproduction event. However, for many species in nature, a large number of offspring are produced simultaneously

in each reproduction event. We call the number of offspring produced (instantaneously) in a single reproduction event the litter size. In general the litter size is stochastic. In this section we describe how to model such a birth processes using a compound Poisson process.

We use $M_t$ to denote the times of reproduction events from 0 to $t$ and we assume that $M_t$ is a Poisson process with intensity $\theta$. And in each reproduction event, the litter size $Z$ obeys a distribution with mean $\mu$ and variance $\sigma^2$. Therefore, the total number of offspring in $[0, t]$ is

$$X(t) = \sum_{k=1}^{M(t)} Z_k. \tag{S3.4}$$

We can derive the mean and variance of $X(t)$.

$$\begin{aligned}
\mathbb{E}X(t) =& \mathbb{E}[\mathbb{E}[X(t)|M(t)]] \\
=& \mathbb{E}[\mathbb{E}[M(t)\mu] \\
=& \theta\mu t, &&\text{(S3.5a)} \\
\mathrm{Var}(X(t)) =& \mathbb{E}[X(t)^2] - (\mathbb{E}[X(t)])^2 \\
=& \mathbb{E}[\mathbb{E}[X(t)^2|M(t)]] - \theta^2\mu^2 t^2 \\
=& \mathbb{E}[M(t)\sigma^2 + M(t)^2\mu^2] - \theta^2\mu^2 t^2 \\
=& \theta(\sigma^2 + \mu^2)t. &&\text{(S3.5b)}
\end{aligned}$$

Thus, $X(t)$ is also a stochastic process with stationary and independent increment which satisfies Assumption S1.1. From Assumption S1.2, we obtain

$$\begin{aligned}
\mathbb{E}(\xi_i) =& B + s\pi_i = \theta_i\mu_i, &&\text{(S3.6a)} \\
\mathrm{Var}(\xi_i) =& \delta_1 B + \delta_2 s\pi_i = \theta_i(\mu_i^2 + \sigma_i^2). &&\text{(S3.6b)}
\end{aligned}$$

Here, $Z$ can be any random variable over non-negative integers, with an arbitrary probability distribution. In this paper we consider two specific cases: $Z$ obeys a Poisson distribution, or a negative binomial distribution.

**Case 1: A Poisson litter size**

First we study the case where $Z$ has the Poisson distribution. For a player with payoff $\pi_i$, we assume that $M_t$ is a Poisson process with parameter $\theta_i$ and $Z$ obeys a Poisson distribution with parameter $\mu_i$. Then, Eq. S3.6 gives that

$$\begin{aligned}
B + s\pi_i =& \theta_i\mu_i, &&\text{(S3.7a)} \\
\delta_1 B + \delta_2 s\pi_i =& \theta_i(\mu_i^2 + \mu_i). &&\text{(S3.7b)}
\end{aligned}$$

The reproduction process of an individual is totally controlled by the two parameters $\theta_i$ and $\mu_i$. On the other hand, the reproducing process is determined by $s\pi_i$. Thus, $\theta_i$ and

9

$\mu_i$ are functions of $s\pi_i$. In the regime of weak selection ($s \ll 1$), any function of $s\pi_i$ can be expand in a Taylor series. Neglecting higher order terms in $s$ leads to $\theta_i$ and $\mu_i$ being linear in $\pi_i$. Thus, we can assume

$$\theta_i = \theta_0 + k_\theta s\pi_i, \tag{S3.8a}$$

$$\mu_i = \mu_0 + k_\mu s\pi_i. \tag{S3.8b}$$

Substituting Eq. S3.8 into Eq. S3.7, we can obtain that

$$B + s\pi_i = \theta_0\mu_0 + (\theta_0 k_\mu + \mu_0 k_\theta)s\pi_i + o(s), \tag{S3.9a}$$

$$\delta_1 B + \delta_2 s\pi_i = \theta_0(\mu_0^2 + \mu_0) + [k_\theta(\mu_0^2 + \mu_0) + \theta_0(2\mu_0 k_\mu + k_\mu)]s\pi_i + o(s). \tag{S3.9b}$$

Comparing the both sides of Eq. S3.9, it yields

$$B = \theta_0\mu_0, \tag{S3.10a}$$

$$1 = \theta_0 k_\mu + \mu_0 k_\theta, \tag{S3.10b}$$

$$\delta_1 B = \theta_0(\mu_0^2 + \mu_0), \tag{S3.10c}$$

$$\delta_2 = k_\theta(\mu_0^2 + \mu_0) + \theta_0(2\mu_0 k_\mu + k_\mu). \tag{S3.10d}$$

There are four equations and four unknowns $(\theta_0, \mu_0, k_\theta, k_\mu)$, and so this system is solvable. We can solve that:

$$\theta_0 = \frac{B}{\delta_1 - 1}, \tag{S3.11a}$$

$$\mu_0 = \delta_1 - 1, \tag{S3.11b}$$

$$k_\theta = \frac{(2\delta_1 - \delta_2) - 1}{(\delta_1 - 1)^2}, \tag{S3.11c}$$

$$k_\mu = \frac{\delta_2 - \delta_1}{B}. \tag{S3.11d}$$

It is worth noting that this construction is meaningful only when $\theta_i$ and $\mu_i$ are positive. Because $s \ll 1$, we expect that $\theta_i$ and $\mu_i$ are positive only when $\theta_0$ and $\mu_0$ are positive, which requires $\delta_1 > 1$. This means that for any parameters $\delta_1$, $\delta_2$, and $B$, provided $\delta_1 > 1$, we can construct such a Poisson-Poisson process that satisfies Assumption S1.2. Note also that a birth process based on Poisson-Poisson process always naturally leads to over-dispersion in the number of offspring.

**Case 2: A negative binomial litter size**

Now we study the case where $Z$ follows a negative binomial Poisson distribution. For a player with payoff $\pi_i$, we assume that $M_t$ is a Poisson process with parameter $\theta_i$, and $Z$ obeys a negative binomial distribution with parameters $q_i$ and $m$ ($q_i \in (0,1)$ and $m \in \mathbb{N}^*$). The negative binomial distribution has mean $q_i m/(1-q_i)$ and variance $q_i m/(1-q_i)^2$. Since

$m$ is an integer and cannot change continuously with $\pi_i$, we assume that all individuals share an identical $m$.

Then, Assumption S1.2 and Eq. S3.5 give

$$B + s\pi_i = \theta_i \left( \frac{q_i m}{1 - q_i} \right), \tag{S3.12a}$$

$$\delta_1 B + \delta_2 s\pi_i = \theta_i \left( \frac{q_i^2 m^2}{(1 - q_i)^2} + \frac{q_i m}{(1 - q_i)^2} \right). \tag{S3.12b}$$

Similarly, we further assume that $\theta_i$ and $q_i$ are linear in $\pi_i$, that is

$$\theta_i = \theta_0 + k_\theta s\pi_i, \tag{S3.13a}$$

$$q_i = q_0 + k_q s\pi_i. \tag{S3.13b}$$

Substituting Eq. S3.13 into Eq. S3.12 and using similar techniques, we can solve to find

$$\theta_0 = \frac{B(1+m)}{(\delta_1 - 1)m}, \tag{S3.14a}$$

$$q_0 = \frac{\delta_1 - 1}{\delta_1 + m}, \tag{S3.14b}$$

$$k_\theta = \frac{(2\delta_1 - \delta_2 - 1)(m+1)}{(\delta_1 - 1)^2 m}, \tag{S3.14c}$$

$$k_q = \frac{(m+1)(\delta_2 - \delta_1)}{B(\delta_1 + m)^2}. \tag{S3.14d}$$

Similarly, this construction is meaningful only when $\theta_0$ is positive and $q_0$ is in $[0, 1]$, which implies $\delta_1 > 1$. And so we see that birth process based on Poisson-negative binomial process always leads to over-dispersion.

Here in our construction, we have assumed $m$ is fixed and the same for all individuals. If $m$ can vary for different individuals, the construction will be more complicated and not unique. Moreover, we have only considered two examples of compound Poisson process here, but many other construction can be derived similarly. And so Assumption S1.2 is very general axiom so that the space of models we analyze covers a wide range of birth processes.

## S4 Extensions

### S4.1 General games

So far, we have analyzed the classic donation games, to explore the effects of demographic stochasticity on the evolution of cooperation. To explore more general results, we extend our model to general two-player/two-action games. A general two-player two-action game has the following payoff structure

$$\begin{array}{c} \\ C \\ D \end{array} \begin{pmatrix} C & D \\ a & b \\ c & d \end{pmatrix}.$$

Here, the two strategies are still called cooperation (C) and defection (D), which are now just generic terms that carry no implied meaning. When two cooperators interact, they both receive payoff $a$. When a cooperator encounters a defector, the cooperator receives $b$ while the defector receives $c$. Mutual defection brings $d$ to both players.

If there are $x$ players employing cooperation and $y$ players employing defection, their payoffs are respectively

$$\pi_C = \frac{xa + yb}{x + y}, \tag{S4.1}$$

$$\pi_D = \frac{xc + yd}{x + y}. \tag{S4.2}$$

As before, we make the parameter transformation $p = x/(x + y)$ and $n = x + y$. Then, under the assumption of $\alpha \gg s$, we can still separate the time-scale of $n$ and $p$, which gives

$$\begin{aligned}\mathrm{d}p =&sp(1 - p)\left[\left(1 - \frac{\delta_2}{M}\right)(b - d + (a - b - c + d)p)\right]\mathrm{d}t \\ &+ \sqrt{\frac{(\delta_1 + 1)Bp(1 - p)}{M}}\left(\sqrt{1 - p}\mathrm{d}W_t^{(1)} - \sqrt{p}\mathrm{d}W_t^{(2)}\right),\end{aligned} \tag{S4.3}$$

where $M$ is the carrying capacity given by $M = \alpha/\lambda$.

### S4.1.1  ODE-based analysis

If the carrying capacity is large, the diffusion term is very small, and thus has little influence on the dynamics. Thus, to provide an rough intuitive interpretation, we first analyze Eq. S4.3 by omitting its diffusion term altogether, which simplifies to an ordinary differential equation (ODE). That is

$$\mathrm{d}p = sp(1 - p)\left[\left(1 - \frac{\delta_2}{M}\right)(b - d + (a - b - c + d)p)\right]\mathrm{d}t. \tag{S4.4}$$

If we ignore $s$ and let $\delta_2 = 0$ in Eq. S4.4, it actually degenerates to the classic replicator equation.

We denote Eq. S4.4 as $\mathrm{d}p = f(p)\mathrm{d}t$ in what follows. This equation has three equilibrium points:

$$p = 0, \tag{S4.5a}$$

$$p = 1, \tag{S4.5b}$$

$$p = p^* = \frac{d - b}{a - b - c + d}. \tag{S4.5c}$$

These equilibrium points are meaningful only if they are located in the domain $[0, 1]$. Games can be classified into three scenarios in terms of the number of equilibrium points within $[0, 1]$ and their stabilities. We will analyze the effects of demographic noise for the three scenarios respectively.

**Dominance games:**

For a dominance game, a player's best choice of strategy does not depend on the opponent's strategy. Without loss of generality, we suppose $a < c$ and $b < d$ such that defection is always the best choice (e.g., prisoner's dilemma).

Since the definition domain of $p$ is $[0,1]$, in this case, $p^* \notin [0,1]$. Thus, Eq. S4.4 has only two fixed points: $p = 0$ and $p = 1$.

If $\delta_2 < M$, we have

$$\left.\frac{\mathrm{d}f(p)}{\mathrm{d}p}\right|_{p=0,\delta_2<M} < 0 \quad \text{and} \quad \left.\frac{\mathrm{d}f(p)}{\mathrm{d}p}\right|_{p=1,\delta_2<M} > 0. \tag{S4.6}$$

Thus, only $p = 0$ (i.e. defectors take over the population) is stable. However, we find that

$$\left.\frac{\mathrm{d}f(p)}{\mathrm{d}p}\right|_{\delta_2<M} \cdot \left.\frac{\mathrm{d}f(p)}{\mathrm{d}p}\right|_{\delta_2>M} < 0, \tag{S4.7}$$

which means $\delta_2 > M$ can alter the stability of equilibrium points. If $\delta_2 > M$, $p = 1$ becomes stable but $p = 0$ is unstable, which implies cooperators will take over the population.

**Coexistence games:**

If $c > a$ and $b > d$, this game is a coexistence game. The best choice for an individual is to choose the opposite strategy of her opponent's. In this case, all of the three points (Eq. S4.5) are in $[0,1]$. Hence the system has three equilibrium points.

When $\delta_2 < M$, we have

$$\left.\frac{\mathrm{d}f(p)}{\mathrm{d}p}\right|_{p=0,\delta_2<M} > 0, \quad \left.\frac{\mathrm{d}f(p)}{\mathrm{d}p}\right|_{p=1,\delta_2<M} > 0, \quad \text{and} \quad \left.\frac{\mathrm{d}f(p)}{\mathrm{d}p}\right|_{p=p^*,\delta_2<M} < 0 \tag{S4.8}$$

Thus, only $p = p^*$ is stable, which means cooperators and defectors will coexist in the equilibrium state.

But for $\delta_2 > M$, $p = 0$ and $p = 1$ becomes stable but $p = (d-b)/(a-b-c+d)$ is unstable, which means cooperators or defectors will finally take over the population. Cooperators and defectors can never coexist.

**Coordination games:**

If $a > c$ and $d > b$, the game is a coordination game. The best choice is to choose the same strategy as your opponent. Similar to the coexistence game, all of the three equilibrium points exist. However, given $\delta_2 < M$, the interior point ($p^*$) is no longer stable and the two points on the boundary ($p = 0$ and $p = 1$) are stable. Thus, the system will eventually converge to full-cooperator or full-defector state according to the initial configuration.

If $\delta_2 > M$, the coordination game can be transformed to a coexistence game. Specifically, $p^*$ becomes stable and another two points are unstable, which means cooperators and defectors will coexist.

13

Generally, under the condition of $\delta_2 > M$, the changes on dynamics can be viewed as doing a transformation on the payoff matrix. For any games, it can transform the payoff matrix to its opposite one, i.e.,

$$\begin{pmatrix} a & b \\ c & d \end{pmatrix} \Rightarrow \begin{pmatrix} -a & -b \\ -c & -d \end{pmatrix}.$$

Thus, compared to the classical outcome, which occurs for $\delta_2 < M$, we have the following effects when $\delta_2 > M$: for a dominance game the dominant strategy is swapped; coexistence games are transformed to coordination games; a coordination games are transformed to coexistence games.

### S4.1.2 Stationary distribution

The analysis above is obtained by omitting the stochastic term, reducing the analysis to an ODE. If we retain the stochastic term, Eq. S4.4 is a Markov process that has two absorbing states $p = 0$ and $p = 1$. Thus, this Markov process is not ergodic and the stationary distribution is not unique. A technical approach to studying this is to assume the boundaries are reflecting, so that no states are absorbing [3]. Mechanistically, this means that when the number of cooperators (defectors) becomes zero, a new cooperator (defectors) arises instantly in the population. Thus the definition domain of $p$ becomes $[1/M, 1 - 1/M]$ approximately. Similar techniques are also used in [6]. This approach makes the Markov process ergodic and the stationary distribution (denoted by $v^*(p)$) is unique. A frequency $p$ with greater density in stationary distribution means that trajectories spend more time in the neighborhood of $p$. The stationary distribution $v^*(p)$ is the solution of the following Fokker-Planck equation:

$$-\frac{\mathrm{d}}{\mathrm{d}p}\mathcal{A}(p)v^*(p) + \frac{1}{2}\frac{\mathrm{d}^2}{\mathrm{d}p^2}\mathcal{B}(p)v^*(p) = 0 \tag{S4.9}$$

where $\mathcal{A}(p)$ and $\mathcal{B}(p)$ is given by

$$\mathcal{A}(p) = sp(1-p)\left(1 - \frac{\delta_2}{M}\right)[b - d + (a - b - c + d)p] \tag{S4.10a}$$

$$\mathcal{B}(p) = \frac{(\delta_1 + 1)Bp(1-p)}{2M} \tag{S4.10b}$$

The solution of this equation can be expressed explicitly [3]:

$$v^*(p) = \frac{\mathcal{N}}{\mathcal{B}(p)} \exp\left[2\int_0^p \frac{\mathcal{A}(p')}{\mathcal{B}(p')}\mathrm{d}p'\right], \tag{S4.11}$$

where $\mathcal{N}$ is a normalization constant such that

$$\int_{1/M}^{1-1/M} v^*(p)\mathrm{d}p = 1. \tag{S4.12}$$

By some basic manipulations, we obtain the stationary distribution. That is

$$v^*(p) = \frac{\mathcal{N}M}{(\delta_1+1)Bp(1-p)} \exp\left[\frac{2sM}{(\delta_1+1)B}\left(1-\frac{\delta_2}{M}\right)\left[(b-d)p + \frac{1}{2}(a-b-c+d)p^2\right]\right] \tag{S4.13}$$

In the following analyse we omit $\mathcal{N}$ since it does not affect the shape of the distribution. We find that Eq. S4.13 is the product of two components:

$$v_1(p) = \frac{M}{(\delta_1+1)Bp(1-p)} \tag{S4.14a}$$

$$v_2(p) = \exp\left[\frac{2sM}{(\delta_1+1)B}\left(1-\frac{\delta_2}{M}\right)\left[(b-d)p + \frac{1}{2}(a-b-c+d)p^2\right]\right]. \tag{S4.14b}$$

Here, $v_1(p)$ is an U-shape distribution, where the boundaries $p = 1$ and $p = 0$ have the largest probability density. For $v_2(p)$, since the exponential function is monotonic increasing, the shape of $v_2(p)$ is determined by the quadratic function inside. If $\delta_2 < M$, for the prisoner's dilemma (dominance game), $v_2(p)$ achieves its extremum at $p = 0$. For a coordination game or a coexistence game, $v_2(p)$ achieves its extremum in the interior ($p = p^*$) or on the boundary ($p = 0$ and $p = 1$) respectively. If $\delta_2 > M$, the maximum of $v_2(p)$ in the case of $\delta_2 < M$ is now turned to be minimum, and vice versa.

if $(\delta_1+1)B$ is small, the shape of $v^*(p)$ is mainly determined by $v_2(p)$ because exponential function is dominant. Thus, we can see the humps of the stationary distribution are in accordance with the stable equilibria in the deterministic analysis. And large $\delta_2$ can alter the position of the humps in stationary distribution, which confirms that large $\delta_2$ can transform the dynamics of a game. However, if $(\delta_1+1)B$ is large, $v^*(p)$ is dominated by $v_1(p)$. The stationary distribution is always U-shape, so all games have similar properties with coordination games. This kind of transformation is not found in ODE-based analysis.

Thus, we can conclude that for a coexistence game, the previously condition we derived for transformation into a coordination game, $\delta_2 > M$, is actually conservative, since $\delta_1$ also works to transform the game into a coordination game. However, for a coordination game, the condition $\delta_2 > M$ is liberal, since the heterogeneity in variance (large $\delta_2$) should offset the effects of $\delta_1$ first when it tries to transform the game into a coexistence game.

### S4.1.3 Games with multiple strategies

All prior analysese have focused on games with only two possible actions. Here we extend our analysis to the case of games with more than two strategies, which includes the famous rock-paper-scissors game with non-transitive payoff structure. For a two-player game with $m$ strategies, suppose its payoff matrix is

$$\mathbf{A} = \begin{pmatrix} a_{11} & \cdots & a_{1m} \\ \vdots & \ddots & \vdots \\ a_{m1} & \cdots & a_{mm} \end{pmatrix}.$$

15

Let $n_i$ denote the number of individuals who adopt strategy $i$ (the population size is $n = \sum n_i$), and $p_i$ denote the frequency of strategy $i$. This gives $p_i = n_i/n$. Collecting all frequencies, we get a vector $\mathbf{p} = [p_1, \cdots, p_n]$. The payoff of individuals with strategy $i$ is

$$\pi_i = (\mathbf{A}\mathbf{p})_i, \tag{S4.15}$$

and its growth equation is

$$\mathrm{d}n_i = n_i[\alpha + s\pi_i - \lambda n]\mathrm{d}t + \sqrt{n_i(\delta_1 B + \delta_2 s\pi_i + D + \lambda n)}\mathrm{d}W_t^{(i)}. \tag{S4.16}$$

Applying Itô's lemma

$$\mathrm{d}p_i = \sum_{k=1}^{m} \frac{\partial p_i}{\partial n_k}\mathrm{d}n_k + \frac{1}{2}\sum_{k=1}^{m} \frac{\partial^2 p_i}{\partial n_k^2}(\mathrm{d}n_k)^2, \tag{S4.17}$$

and separating the time-scale of $n$ and $p$ (assuming $\alpha \gg s$), we obtain the stochastic differential equation for $p_i$:

$$\mathrm{d}p_i = s\left(1 - \frac{\delta_2}{M}\right)p_i\left(\pi_i - \bar{\pi}\right)\mathrm{d}t + \sqrt{\frac{(\delta_1+1)Bp_i}{M}}\mathrm{d}W_t^{(i)} - p_i\sum_{j=1}^{m}\sqrt{\frac{(\delta_1+1)Bp_j}{M}}\mathrm{d}W_t^{(j)} \tag{S4.18}$$

where $\bar{\pi} = \sum p_i\pi_i$ is the average payoff. This is a $n-1$ dimensional system. For large $M$, the diffusion term can also be ignored. This shows that for games with multiple strategies, when $\delta_2 > M$, the sign of the drift term can also be changed. And so once again, similar to the case of two-action games, we find that large heterogeneity in offspring variance due to payoff (large $\delta_2$) can reverse the direction of evolution.

Using similar techniques, we can also analyze the stationary distribution with reflecting boundaries. Suppose the stationary distribution is $v^*(\mathbf{p}) = v^*(p_1, p_2, \cdots, p_{n-1})$. The corresponding Fokker-Planck equation is

$$-\sum_{i=1}^{n-1}\frac{\partial}{\partial p_i}\mathcal{A}_i(\mathbf{p})v^*(\mathbf{p}) + \frac{1}{2}\sum_{i,j=1}^{n-1}\frac{\partial^2}{\partial p_i\partial p_j}\mathcal{B}_{ij}(\mathbf{p})v^*(\mathbf{p}) = 0, \tag{S4.19}$$

where

$$A_i(\mathbf{p}) = \left(s - \frac{s\delta_2}{M}\right)p_i(\pi_i - \bar{\pi}) \tag{S4.20}$$

$$B_{ij}(\mathbf{p}) = \begin{cases} -\frac{(\delta_1+1)B}{M}p_ip_j & i \neq j \\ \frac{(\delta_1+1)B}{M}p_i(1-p_i) & i = j \end{cases}. \tag{S4.21}$$

Unfortunately, it is difficult to obtain an explicit solution for this equation. And so we resort to simulations to show that large offspring variance can qualitatively change the directions of evolution and the stationary distribution (Fig. S4).

16

## S4.2 Dynamics in slow-growing populations

In all our analyses above we have assumed that the net baseline growth rate, $\alpha = B - D$, is sufficiently large compared to selection intensity $s$, so that the population rapidly approaches carrying capacity. In this setting the analysis can be simplified to a one-dimensional system (for two-strategy games), by separation of timescales. By contrast, in this section we focus on the case in which the net baseline growth rate of the population is small. In particular, we study the case when baseline growth has the same order as selection, $\alpha = O(s)$. In this regime the population grows very slowly and fixation of one type or another typically occurs long before reaching carrying capacity. The key question still remains: can large offspring variance qualitatively change the evolutionary outcome? We analyze this question in the case of the donation game.

### S4.2.1 Fixation probability

Since the system can no longer be simplified to a single dimension along the slow manifold, the fixation probability will now depend on the initial population size as well as the initial frequency of cooperators. Suppose there are $n_0$ individuals and a portion $p_0$ of them are cooperators initially. Given Eq. S2.2, the fixation probability $\rho(p_0, n_0)$ of cooperators satisfies the following backward Kolmogorov equation (in the following derivation, we omit the subscripts and write the fixation probability as $\rho(p, n)$):

$$\begin{cases} \mathcal{G}\rho(p, n) = 0, \\ \rho(1, n) = 1, \quad n > 0 \\ \rho(0, n) = 0, \quad n > 0 \end{cases} \tag{S4.22}$$

where

$$\mathcal{G}f = \left[ -scp(1-p) + \delta_2 sc \frac{p(1-p)}{n} \right] \frac{\partial f}{\partial p} + [\alpha + s(b-c)pn - \lambda n^2] \frac{\partial f}{\partial n} \\ + \frac{(\delta_1 B + D + \lambda n)p(1-p)}{2n} \frac{\partial^2 f}{\partial p^2} + \frac{(\delta_1 B + D + \lambda n)n}{2} \frac{\partial^2 f}{\partial n^2} \tag{S4.23a}$$

is the infinitesimal generator of Eq. S2.2. In the neutral case ($s = 0$), we can verify that the solution of Eq. S4.22 is $\rho = p$. Thus, the fixation probability equals the initial frequency $p_0$ when $s = 0$. For weak selection, inspired by the form of fixation probability in Eq. S2.9 and the boundary conditions of Eq. S4.22, we make the ansatz that the solution has the following form [5]

$$\rho(p, n) = p + sp(1-p)\phi(n) + o(s). \tag{S4.24}$$

Substituting Eq. S4.24 into Eq. S4.22 and only retaining the first-order term of $s$ (note that $\alpha = O(s)$ and $\lambda \ll s$ due to the large carrying capacity), we can obtain

$$-c + \frac{\delta_2 c}{n} - \frac{\delta_1 B + D}{n}\phi + \frac{(\delta_1 B + D)n}{2}\phi'' = 0. \tag{S4.25}$$

17

This is an Euler-Cauchy equation, whose solution is

$$\phi(n) = A_1 n^2 + \frac{A_2}{n} - \frac{cn}{\delta_1 B + D} + \frac{\delta_2 c}{\delta_1 B + D}. \tag{S4.26}$$

where $A_1$ and $A_2$ are arbitrary constants. Substitution into Eq. S4.24 yields

$$\rho(p,n) = p + sp(1-p)\left(A_1 n^2 + \frac{A_2}{n} - \frac{cn}{\delta_1 B + D} + \frac{\delta_2 c}{\delta_1 B + D}.\right) \tag{S4.27}$$

To satisfy the boundary conditions (i.e. $\rho(0,n) = 0$ and $\rho(1,n) = 1$), $p(1-p)\phi(n)$ must equal to 0 when $p = 0$ and $p = 1$, which has already been satisfied. Furthermore, when $c = 0$, cooperators and defectors always have the same payoffs. Thus, in this case, the evolution is also equivalent to a neutral drift and the fixation probability must equal $p$, which yields $p(1-p)\phi(n) = 0$ for all $p$ when $c = 0$. We obtain $A_1 = A_2 = 0$. So the fixation probability of cooperators is

$$\rho = p_0 - \frac{sc}{\delta_1 B + D} p_0 (1 - p_0)(n_0 - \delta_2). \tag{S4.28}$$

Note that this expression for the fixation probability is similar to Eq. S2.9 (ignoring the difference in the coefficient), but the carrying capacity $M$ has been replaced by initial population size $n_0$. If

$$\delta_2 > n_0, \tag{S4.29}$$

then the fixation probability exceeds $p_0$, which means cooperation is be favored by selection. The form of this condition is similar to Eq. S2.10. Moreover, numerical simulations and individual-based simulations (based on compound Poisson processes) both verify that this condition accurately predicts when cooperation will be favored in a slow-growing population (Fig. S5 and S6).

### S4.2.2 Extinction and persistence

When $\alpha \sim s$ the population size grows slowly. For small initial population size $n_0$, fixation occurs rapidly, so that fixation typically occurs before the population reaches carrying capacity. After cooperators or defectors become fixed, the system then becomes a one-dimensional process, where all individuals have the same payoff and thus have the same variance in offspring number. The population will then either go extinct, or grow logistically until the population reaches the carry capacity (Fig. S7) (and thereafter persist for exponentially long in the carrying capacity). Here, we analyze the process of extinction or persistence in the case when cooperators fix (the procedures are analogous in case of defectors fixing first).

If cooperators become fixed, the system then contains only cooperators for all subsequent evolution. Therefore it becomes a one-dimensional diffusion process, which is given by

$$\mathrm{d}x = x[\alpha + s(b - c) - \lambda x]dt + \sqrt{x(\delta_1 B + D + \lambda x)}\mathrm{d}W_t^{(1)}. \tag{S4.30}$$

By analyzing the corresponding deterministic system ($\delta_1 = \delta_2 = 0$), the carrying capacity of cooperators is given by

$$M_C = \frac{\alpha + s(b-c)}{\lambda}. \tag{S4.31}$$

For the stochastic system, we first analyze the case of no competition for resources (i.e., $\lambda = 0$, and so no carrying capacity). The equation becomes

$$\mathrm{d}x = x[\alpha + s(b-c)]dt + \sqrt{x(\delta_1 B + D)}\mathrm{d}W_t^{(1)}. \tag{S4.32}$$

When we ignore the stochastic term, this equation represents exponential growth. The population size grows without bound. When the stochastic term is considered, given there are $x_f$ cooperators after they become fixed, there are two scenarios: grow to infinity or go extinct, since $x = 0$ is an absorbing state. The probability that cooperators go extinct given that there are $x_f$ cooperators after their fixation can be computed by [3]

$$P_C = \frac{\int_{x_f}^{\infty} s(z)\mathrm{d}z}{\int_0^{\infty} s(z)\mathrm{d}z}. \tag{S4.33}$$

where

$$s(z) = \exp\left(\int -\frac{2E(z)}{D(z)}\mathrm{d}z\right), E(z) = z[\alpha + s(b-c)], D(z) = \delta_1 Bz + Dz. \tag{S4.34}$$

After some manipulations, we obtain

$$P_C = \exp\left(-\frac{2(s(b-c) + \alpha)x_f}{\delta_1 B + D}\right), \tag{S4.35}$$

which is termed as perishing probability here. And the probability of growing to infinity is thus $1 - P_C$.

When competition for resources is considered ($\lambda \neq 0$), the population can never grow without bound. There are only one absorbing state $x = 0$ and the population size is bounded due to the restriction of carrying capacity. Thus, for all $x_f$, the probability that cooperators finally perish is 1. However, the population usually does not perish directly after they fix, but the population still grows to the carrying capacity and fluctuates around it for a very long time (in fact, exponentially long in the carrying capacity) before their extinction (Fig. S7). Thus, when the carrying capacity is large, in any practical application the population will not go extinct once it has reached carrying capacity. So we can artificially divide the dynamics into two scenarios: given there are $x_f$ cooperators after they become fixed, the cooperators will either perish directly (Fig. S7A), or they grow to carrying capacity and then fluctuate around it for an extremely long time (Fig. S7C). Since the initial population size is small, the competition among individuals can be ignored before the population size reaches the carrying capacity. The probability of perishing directly can be approximated by the case of no competition, i.e. $P_C$. And the probability of reaching carrying capacity is approximated by $1 - P_C$. This approximation agrees well with simulations (Fig. S8A).

Similarly, if defectors become fixed, then the extinction probability can also be obtained using the same method. That is

$$P_D = \exp\left(-\frac{y_f \alpha}{\delta_1 B + D}\right). \tag{S4.36}$$

And if they do not go extinct rapidly they will grow and finally reach their carrying capacity, which is given by

$$M_D = \frac{\alpha}{\lambda}. \tag{S4.37}$$

Note that the carrying capacities for cooperators and defectors are no longer identical compared to the case of $\alpha \gg s$. The cooperators have an extra advantage in that their carrying capacity is larger than defectors'.

Now, we consider the two processes (fixation and extinction/persistence) together. Suppose there are $x_0$ cooperators and $y_0$ defectors in the population initially. We call the probability that cooperators become fixed and then keep growing to carrying capacity $M_C$ the "persistence probability" (for cooperators). Thus, the persistence probability is determined by the combination of the fixation probability and (one minus) the extinction probability. However, the exact number of cooperators or defectors when they become fixed (i.e. $x_f$ and $y_f$) is difficult to analyze. So it is difficult to give an explicit expression for the persistence probability. However, intuitively, since fixation is fast and the growth rate $\alpha$ is small, the population size after fixation will not change much compared to the initial population size. Thus, we expect that $x_f$ and $y_f$ have the same order as $n_0$.

We have already shown that when $\delta_2 > M$, cooperation will be favored by the selection. Here, we show that even when $\delta_2 < M$, cooperators may have a greater persistence advantages than defectors. By Eq. S4.28, we see that when $\delta_2 < M$, larger variance in due to baseline births $\delta_1$ is beneficial to the fixation of cooperators, but it also leads to more likely extinction after they fix (see Eq. S4.35). Thus we predict that intermediate $\delta_1$ is most beneficial for cooperators' persistence. For defectors, larger $\delta_1$ is detrimental to their fixation and also detrimental to their growth to $M_D$ after fixation. Thus, large $\delta_1$ is always detrimental to defectors persisting (Fig. S8B). For some intermediate $\delta_1$, cooperators may have greater persistence probability than defectors, which is a different sense in which offspring variance can foster the evolution of cooperation.

Similarly, we find that given a fixed initial frequencey $p_0$, larger initial population size $n_0$ is detrimental to cooperators' fixation but beneficial for their growth to carrying capacity $M_C$ after fixation (since $x_f \sim n_0$). And so we also predict that an intermediate value of the initial population size $n_0$ is mosts beneficial to cooperators' persistence. But for defectors, larger $n_0$ is always beneficial for defector's persistence (Fig. S8C).

Based on the discussions above, we conclude that cooperators may have an advantage over defectors in stochastic populations for three distinct reasons: (1) lager fixation probability; (2) larger carrying capacity; (3) larger persistence probability.

# Appendix: Details of simulations

Two methods of simulation are used in this work: numerical simulations by sampling the SDE, and Monte Carlo simulations based on compound Poisson process. Here, we provide more details about these simulations.

### Numerical simulations:

Numerical simulations are based on Eq. S1.8. We choose a small time step $\Delta t = 0.01$ and set the initial configuration ($x_0$ and $y_0$). The Euler-Maruyama scheme is used to sample Eq. S1.8 [7], which is given by

$$x_{t+\Delta t} = x_t + x_t(\alpha + s\pi_C - \lambda(x_t + y_t))\Delta t + \sqrt{x_t(\delta_1 B + \delta_2 s\pi_C + D + \lambda(x_t + y_t))}\Delta W_t^{(1)} \tag{S4.38a}$$

$$y_{t+\Delta t} = y_t + y_t(\alpha + s\pi_D - \lambda(x_t + y_t))\Delta t + \sqrt{y_t(\delta_1 B + \delta_2 s\pi_D + D + \lambda(x_t + y_t))}\Delta W_t^{(2)}. \tag{S4.38b}$$

Here, $\Delta W_t^{(1)} \sim N(0, \Delta t)$ and $\Delta W_t^{(2)} \sim N(0, \Delta t)$ are two independent Gaussian random variables. In each time step, by sampling random numbers from $\Delta W_t^{(1)}$ and $\Delta W_t^{(2)}$, we can obtain the population composition in the next time step. By iteration of this procedure, a trajectory of evolution is obtained. Then, we can simulate a large number replicate trajectories and compute the fixation probability.

### Monte Carlo simulations:

Monte Carlo simulations must rely on a explicit reproducing process. Here, as we showed in Section S3, we assume the birth process is a compound Poisson process and the death process is a classic Poisson process.

For compound Poisson process, the times of birth events is still a classic Poisson process. Thus, we can consider the birth events and death events together. We refer to them collectively as the updating events. In the following, we only illustrate the procedure of simulation by taking the Poisson-Poisson process as an example, which is similar to Poisson-Negative binomial process.

For a cooperator, its birth event takes place at rate $\theta_C$ (i.e., the parameter of $M_t$ in Section S3), and $\theta_D$ for a defector. In each birth event, the litter size obeys a Poisson distribution with parameter $\mu_C$ ($\mu_D$ for a defector). Using the property of Poisson process that the sum of two Poisson process is also a Poisson process with rate summed, the birth events of all cooperators take place at rate $x\theta_C$, and $y\theta_D$ for defectors. Similarly, the death events take place at rate $Dx + \lambda x(x+y)$ for cooperators, and $Dy + \lambda y(x+y)$ for defectors. Thus, the updating events take places with rate

$$R = x\theta_C + y\theta_D + Dx + \lambda x(x+y) + Dy + \lambda y(x+y). \tag{S4.39}$$

In the simulation, we first set the initial configuration of population ($x_0$ and $y_0$). Suppose

in time $T$, there are $x_T$ cooperators and $y_T$ defectors. Since the time interval between two consecutive events of a Poisson process follows an exponential distribution with the same parameter of the Poisson process, we sample a value of $\Delta T$ according to an exponential distribution with parameter $R$. Then, in time $T + \Delta T$, the population composition is computed by the following rule:

**Step 1.** We generate a random number $r$ which is uniformly sampled from $[0, 1]$.

**Step 2.** The population composition updates in one of the following ways according to the value of $r$.

- Cooperators' birth: If $r < x\theta_C/R$ (with probability $x\theta_C/R$), then $x_{T+\Delta T} = x_T + \delta x$ and $y_{T+\Delta T} = y_T$, where $\delta x$ is an integer randomly sampled from a Poisson distribution with parameter $\mu_C$ (Poisson-Poisson process).

- Defectors' birth: If $x\theta_C/R \leq r < (x\theta_C + y\theta_D)/R$ (with probability $y\theta_C/R$), then $x_{T+\Delta T} = x_T$ and $y_{T+\Delta T} = y_T + \delta y$, where $\delta y$ is an integer randomly sampled from a Poisson distribution with parameter $\mu_D$ (Poisson-Poisson process).

- Cooperators' death: If $(x\theta_C + y\theta_D)/R \leq r < (x\theta_C + y\theta_D + Dx + \lambda x(x+y))/R$ (with probability $(Dx + \lambda x(x+y))/R$), then $x_{T+\Delta T} = x_T - 1$ and $y_{T+\Delta T} = y_T$.

- Defectors' death: If $((x\theta_C + y\theta_D + Dx + \lambda x(x+y))/R \leq r < 1$ (with probability $(Dy + \lambda y(x+y))/R$), then $x_{T+\Delta T} = x_T$ and $y_{T+\Delta T} = y_T - 1$.

Thus, we obtain the population composition in time $T + \Delta T$. Next, we sample another $\Delta T$, and the algorithm enters the next cycle. Recording all time steps and the corresponding population composition, we obtain a trajectory of evolution.

# Supplementary figures

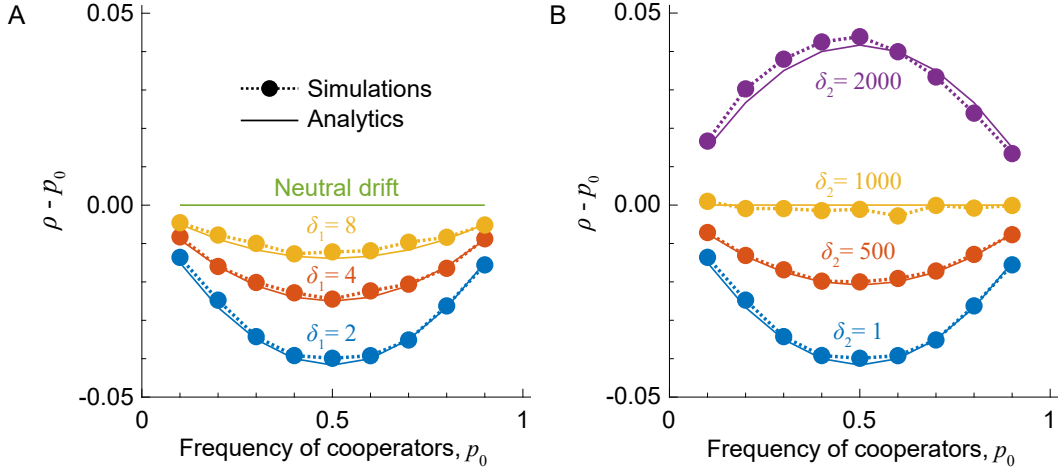

**Figure S2. The initial frequency of cooperators and $\delta_1$ can affect the fixation probability.** Under neutral drift, the fixation probability equals to the initial frequency of cooperators, $p_0$. The panels show the fixation probabilities minus $p_0$. If $\rho - p_0$ exceeds zero cooperation is favored by natural selection, and vice versa. (A) For all $\delta_1$, cooperation is never favored by selection, although large $\delta_1$ makes the fixation probability closer to neutral drift. Furthermore, the fixation probability for intermediate initial frequency $p_0$ deviates the most from neutrality. (B) By contrast, large $\delta_2$ can cause selection to favor fixation of cooperators. For the case of $M = 1,000$, when $\delta_2 < 1,000$, cooperation is never favored and $p = 0.5$ is the most detrimental to the evolution of cooperation. But when $\delta_2 > M$, all fixation probabilities exceed that of neutral drift, and $p_0 = 0.5$ is the most beneficial to cooperation. And when $\delta_2 = M$, the fixation probability equals to the neutral case. Numerical simulations (dots) agree well with analytical approximations (dashed lines). Parameters: $B = 2$, $D = 1$, $b = 1.1$, $c = 1$, $s = 0.001$, $n_0 = 500$, $\lambda = 10^{-3} (M = 1000)$, $\delta_2 = 1$ (A), $\delta_1 = 2$ (B).

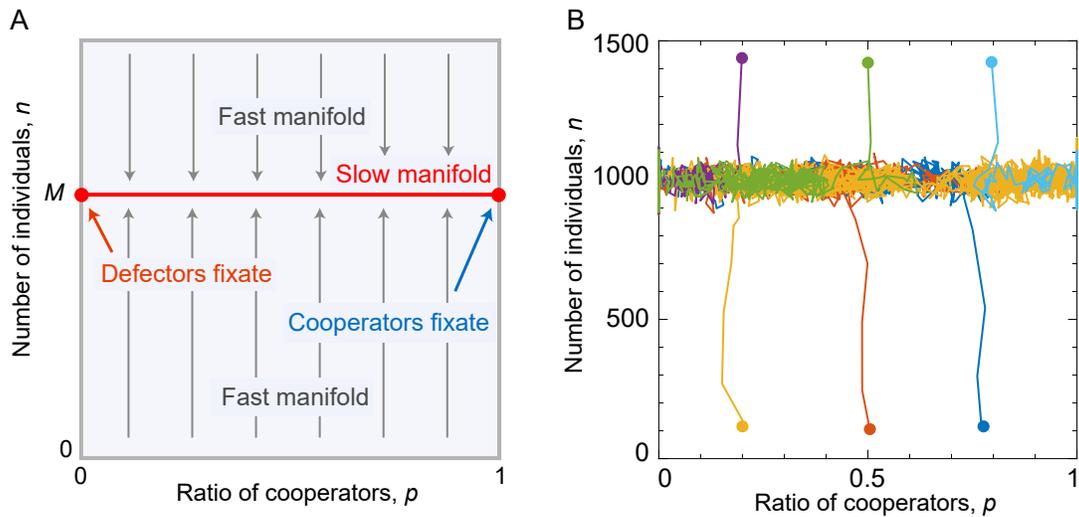

**Figure S3. Simulations showing fast and slow manifolds.** (A) When the growth rate is much larger than selection intensity ($\alpha \gg s$), population size equilibrates much faster than the frequency of cooperators: $n$ is the fast variable and $p$ is the slow variable. In a deterministic analysis, all trajectories rapidly approach the slow manifold, without changes in $p$, and then move along the slow manifold. (B) Simulations of the stochastic model illustrate the time-scale separation. For six different settings of initial states (solide point), we show trajectories produced by numerical simulation of Eq. S1.8. The trajectories exhibit behavior that agrees well with the deterministic analysis of the fast-slow system. Parameters: $B = 2$, $D = 1$, $\lambda = 0.001$, $\delta_1 = \delta_2 = 1$, $b = 1.1$, $c = 1$, $s = 0.001$.

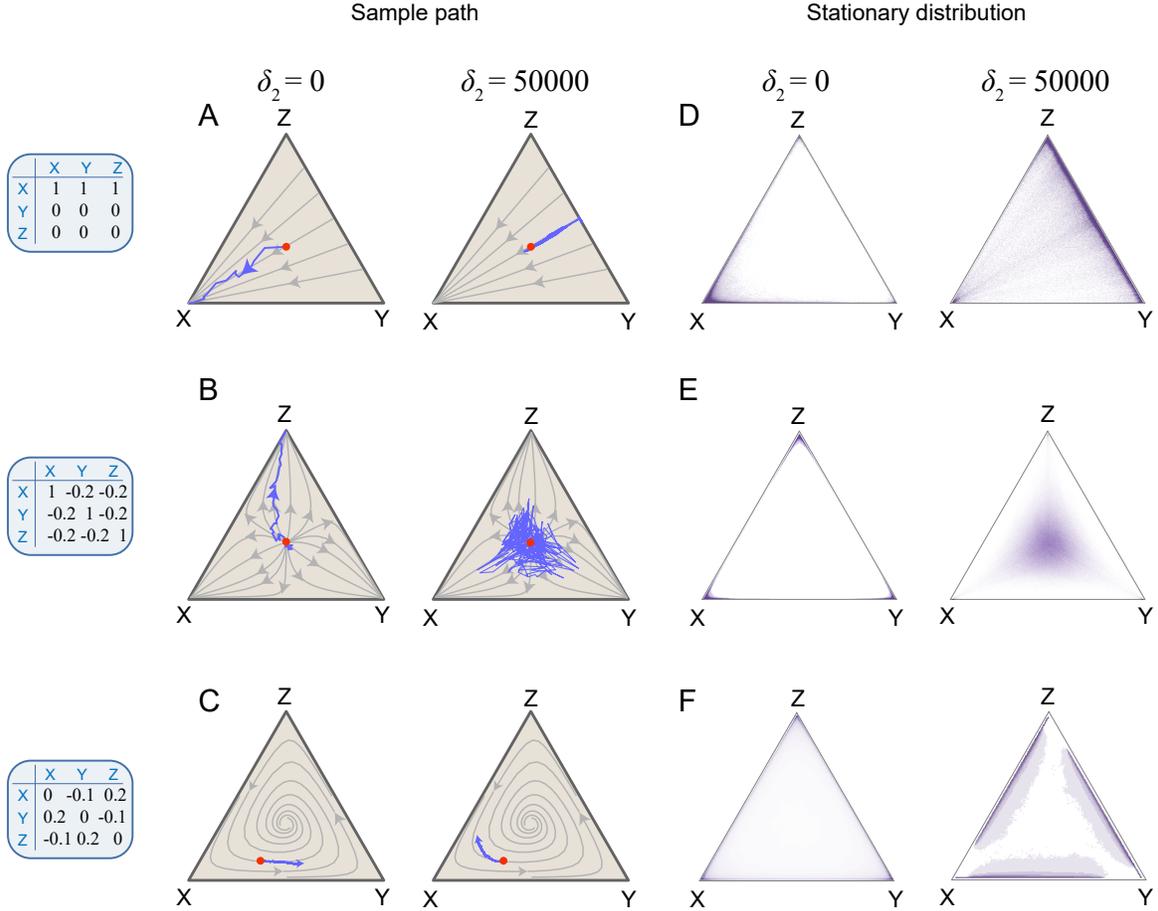

Figure S4. **Demographic noise produces qualitatively different evolutionary outcomes in a game with three strategies.** We show evolutionary trajectories in simplices representing three strategic frequencies in three different games (rows A, B, C). Blue trajectories show simulations of the stochastic system starting from the red point; and gray trajectories show the dynamics predicted by the classic replicator dynamics in an infinite population. When the offspring variance related to payoff, $\delta_2$, is sufficiently large, it can convert an unstable (stable) equilibrium point into a stable (unstable) equilibrium (A, B). For the rock-paper-scissors game (row C), the classic replicator dynamics predicts a spiral sink towards a stable point with all three strategies present. The stochastic system does not follow any of these trajectories exactly, but its direction can be reversed when $\delta_2$ is large (panel C shows the ensemble average trajectory over 1000 simulations). Furthermore, if we assume that the boundary is reflecting, the stationary distributions also show that large $\delta_2$ qualitatively changes the qualitatively (D, E, F). Parameters: $B = 2$, $D = 1$, $s = 0.001$, $\delta_1 = 2$, $\lambda = 10^{-4}$.

25

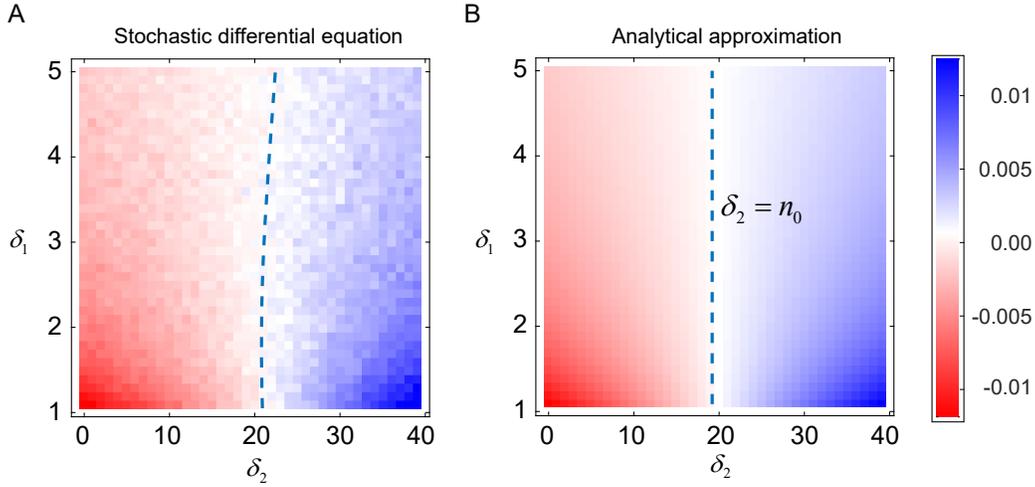

**Figure S5. Cooperation is favored if $\delta_2 > n_0$ in a slow-growing population.** When the baseline growth rate is small, namely $\alpha \sim s$, we can still obtain an analytic prediction for the fixation probability of cooperation. In this figure, each pixel represents the fixation probability minus $p_0$ (the neutral fixation probability). The blue region means cooperation is favored and the red region means defection is favored. (A) shows the fixation probabilities sampled from the stochastic differential equation (Eq. S1.8). (B) is obtained from our analytical approximation (Eq. S4.28). As predicted, when $\delta_2$ exceeds the initial population size, namely, $\delta_2 > n_0$, cooperators are favored by selection in a slow-growing population. Parameters: $b = 1.1$, $c = 1$, $s = 0.005$, $\lambda = 10^{-5}$, $B = 2$, $D = 1.995$.

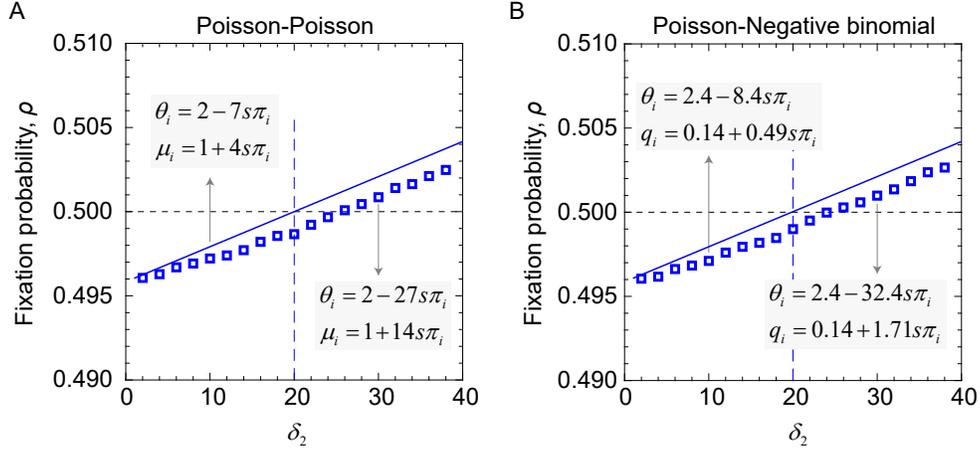

**Figure S6. Selection for cooperation in a compound Poisson process, for a slow-growing population.** Here, we consider the case of a slow-growing population, with $\alpha \sim s$. We simulated a compound Poisson birth process with either a Poisson-distributed litter size (A) or a negative binomial litter size (B). The parameters of the birth process ($\theta_i$ and $\mu_i$ in panel A; $\theta_i$ and $q_i$ in panel B) can be chosen to satisfy our general conditions for the mean and variance in total offspring produced per unit time, for any choice of $\delta_1 > 1$, $\delta_2$, and $B$. Two examples with the parameters that correspond to $(\delta_1 = 2, \delta_2 = 10)$ and $(\delta_1 = 2, \delta_2 = 30)$ are shown in each panel. Blue squares indicate the fixation probability of cooperators, starting from an initial population with $x_0 = y_0 = 10$ ($n_0 = 20$), observed in $10^7$ replicate Monte Carlo simulations. Selection favors cooperation if the fixation probability $\rho$ exceeds the initial fraction of cooperators, $p_0 = 0.5$ (horizontal dashed line). The solid lines plot our analytical approximation for the fixation probability (Eq. S4.28). As predicted by our analysis, cooperation is favored when $\delta_2 > n_0$ in a slow-growing population. Parameters: $B = 2$, $D = 1.998$, $s = 0.005$, $b = 1.1$, $c = 1$, $\lambda = 10^{-6}$, $\delta_1 = 2$, $m = 5(B)$.

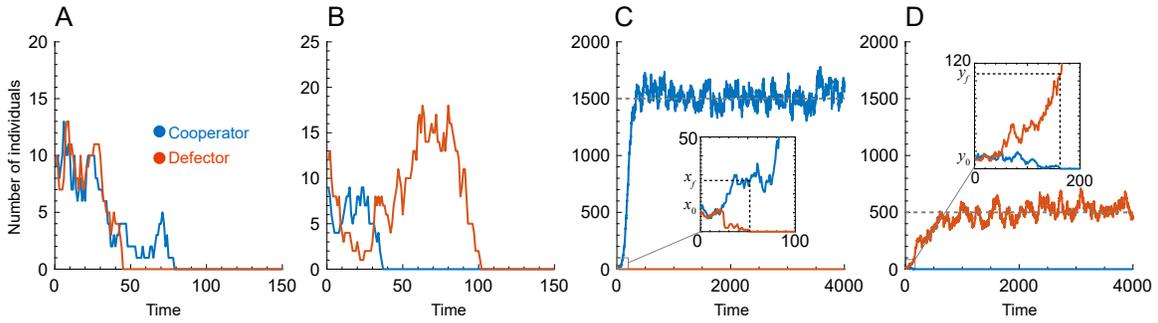

**Figure S7. Sample trajectories of evolution in a slow-growing population.** If the baseline growth rate is small ($\alpha \sim s$), given a small initial population then fixation always occurs before the population size reaches carrying capacity. Since there is only one absorbing state ($x = y = 0$), all trajectories eventually absorb into extinction. However, if one phenotype reaches carrying capacity, the time to absorption is exponentially long in the carrying capacity, so that the population effectively persists for large finite times. We therefor classify the trajectories into four types: (A) Cooperators fix but then perish. (B) defectors fix but then perish. (C) Cooperators fix and then reach their carrying capacity $M_C$. (D) Defectors fix and then reach their carrying capacity $M_D$. Parameters: $b = 3$, $c = 1$, $s = 0.01$, $B = 0.25$, $D = 0.24$, $x(0) = y(0) = 10$, $\lambda = 2 \times 10^{-5}$, $\delta_1 = \delta_2 = 1$.

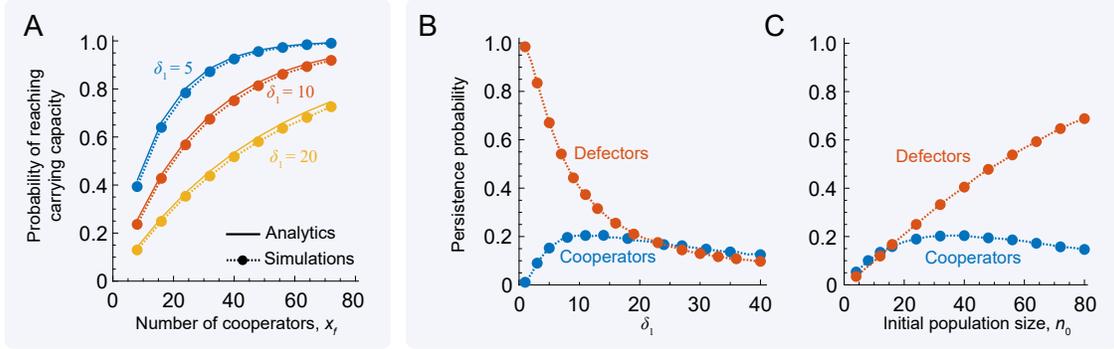

**Figure S8. Persistence probabilities of cooperators and defectors in slow-growing populations.** When the baseline growth rate is small (i.e., $\alpha \sim s$), fixation of one type or the other will occur before the population reaches carrying capacity. The probability that a phenotype not only fixes, but also reaches its carrying capacity in subsequent evolution is called the persistence probability. (A) Take cooperators as an example. If there are $x_f$ cooperators when they become fixed, the probability that they grow to carrying capacity can be estimated by the probability that the population grows without bound in the case without resource competition. This approximation agrees well with simulations. (B) For cooperators, values of $\delta_1$ that are either too large or too small are both detrimental to cooperator persistence. Intermediate values of $\delta_1$ are most beneficial to cooperator persistence. For defectors, larger $\delta_1$ is always detrimental to defector persistence. (C) Similarly, Intermediate values of the initial population size is most beneficial to cooperator persistence. However, larger initial population sizes are always beneficial for defector persistence. These results agree with our theoretical analysis in Section S4.2.2. Parameters: $b = 2$, $c = 1$, $s = 0.01$, $B = 0.1$, $D = 0.09$, $\lambda = 10^{-5}$, $\delta_2 = 1$, $\delta_1 = 10$ (C), $x_0 = y_0 = 20$ (B), $p_0 = 0.5$ (C).

29


\end{document}